\documentclass[twocolumn]{aastex62}

\usepackage{color}
\newcommand{\kawashima}[1]{\textcolor{black}{#1}}

\received{}
\revised{}
\accepted{}
\submitjournal{AJ}

\shorttitle{Sample article}
\shortauthors{Kawashima \& Rugheimer}
\begin{document}

\title{Theoretical Reflectance Spectra of Earth-Like Planets through Their Evolutions: \\
Impact of Clouds on the Detectability of Oxygen, Water, and Methane with Future Direct Imaging Missions}

\correspondingauthor{Yui Kawashima}
\email{y.kawashima@sron.nl}

\author[0000-0003-3800-7518]{Yui Kawashima}
\affiliation{SRON Netherlands Institute for Space Research, Sorbonnelaan 2, 3584 CA Utrecht, The Netherlands}
\affiliation{Earth-Life Science Institute, Tokyo Institute of Technology 2-12-1-IE-1 Ookayama, Meguro-ku, Tokyo 152-8550, Japan}
\affiliation{Department of Earth and Planetary Science, Graduate School of Science, The University of Tokyo, 7-3-1 Hongo, Bunkyo-ku, Tokyo 113-0033, Japan}

\author{Sarah Rugheimer}
\affiliation{Atmospheric, Oceanic, and Planetary Physics Dept, University of Oxford, Clarendon Laboratory, Parks Road, Oxford OX1 3PU, UK}
\affiliation{School of Earth and Environmental Science, University of St Andrews, Irvine Building, St. Andrews KY16 9AL, UK}



\begin{abstract}

In the near-future, atmospheric characterization of Earth-like planets in the habitable zone will become possible via reflectance spectroscopy with future telescopes such as the proposed LUVOIR and HabEx missions.
While previous studies have considered the effect of clouds on the reflectance spectra of Earth-like planets, the molecular detectability considering a wide range of cloud properties has not been previously explored in detail.
In this study, we explore the effect of cloud altitude and coverage on the reflectance spectra of Earth-like planets at different geological epochs and examine the detectability of $\mathrm{O_2}$, $\mathrm{H_2O}$, and $\mathrm{CH_4}$ with test parameters for the future mission concept, LUVOIR, using a coronagraph noise simulator previously designed for WFIRST-AFTA.
Considering an Earth-like planet located at 5~pc away, 
%we find that for the proposed LUVOIR telescope, the $\mathrm{O_2}$ A-band feature (0.76~$\mu$m) will be detectable with less than 100-hour integration time for a modern Earth-like atmosphere unless high coverage ($\gtrsim 80$\%) cloud exists at high altitudes ($\gtrsim 16$~km).
%For the atmosphere with 10\% of modern Earth $\mathrm{O_2}$ abundance, its detection with $\lesssim$~100~hours will be possible as long as there are predominantly lower altitude ($\lesssim 10$~km) clouds with a global cloud coverage of $\gtrsim 20\%$.
%For the 1\% of modern Earth $\mathrm{O_2}$ abundance case, its detection with less than 100~hours is impossible.
we have found that for the proposed LUVOIR telescope, the detection of the $\mathrm{O_2}$ A-band feature (0.76~$\mathrm{\mu}$m) will take approximately 100, 30, and 10~hours for the majority of the cloud parameter space modeled for the atmospheres with 10\%, 50\%, and 100\% of modern Earth O$_2$ abundances, respectively.
Especially, for \kawashima{the case of $\gtrsim 50$\%} of modern Earth O$_2$ abundance, the feature will be detectable with integration time $\lesssim 10$~hours as long as there are lower altitude ($\lesssim 8$~km) clouds with a global coverage of $\gtrsim 20\%$.
For the 1\% of modern Earth $\mathrm{O_2}$ abundance case, however, it will take more than 100~hours for all the cloud parameters we modeled.
%\kawashima{However, detecting the $\mathrm{O_2}$ A-band for $\mathrm{O_2}$ abundances of $< 1$\% of the modern value would require integration times of greater than 100~hours.}
\end{abstract}



\section{Introduction} \label{sec:intro}
With recent advances in observational techniques, more than 3000 exoplanets have been reported so far\footnote{http://exoplanets.org} with many more nearby habitable exoplanets expected to be discovered by TESS.
Already, some rocky planets have been found in habitable zones (HZs) of their host stars such as Proxima Centauri~b, TRAPPIST-1 (e, f, and g), and LHS~1140b \citep{2016Natur.536..437A, 2017Natur.542..456G, 2017Natur.544..333D}.
The next step will be to characterize the atmospheres of these planets.
For characterization of planets in the habitable zones, reflectance spectroscopy is most suitable for the planets around F, G, and K-type stars because of the larger angular separation of the HZs from those host stars.
Transmission spectroscopy suits the characterization of the planets in the HZs around M dwarfs because of their larger transit probabilities and larger planet-to-star radius ratios.

The first telescopes capable of characterizing rocky habitable planet's atmospheres will be JWST (launching in 2021) and through high-resolution spectroscopy with large ground-based telescopes coming online in the 2020s such as ELT (39~m).
However, these missions will only be able to characterize a handful of habitable worlds.
As such, future mission concepts like LUVOIR and HabEx are being proposed that would be able to detect and characterize statistically meaningful samples \citep[see][]{2014ApJ...795..122S, 2015ApJ...808..149S}.
Compared to JWST \kawashima{with the diameter of 6.5~m and the wavelength coverage of 0.6-28.5~$\mu$m}, LUVOIR is proposed to have a much larger diameter of 15 or 8~m and would probe shorter wavelength \kawashima{range} of 0.1-2.5~$\mathrm{\mu}$m and a coronagraph with the possibility of a starshade.\footnote{https://asd.gsfc.nasa.gov/luvoir/} HabEx, a 4~m telescope, is proposed to have a starshade and a coronagraph and likewise will probe shorter wavelength \kawashima{range} than JWST, 0.2-1.8~$\mathrm{\mu}$m.\footnote{https://www.jpl.nasa.gov/habex/} LUVOIR and HabEx will be suitable for the detection and characterization of planets in the HZs around F, G, and K-type stars via reflectance spectroscopy, while JWST is best suited for transiting planets in the HZs around M dwarfs.

Among the several proposed biosignature gases, the existence of molecular oxygen in the atmosphere has been long considered as one of the most promising biosignature candidates for Earth-like planets \citep[see reviews by][and references therein]{2017AsBio..17.1022M, 2018AsBio..18..630M}.
Although several abiotic sources of $\mathrm{O_2}$ have been proposed so far
\citep{2012ApJ...761..166H, 2014E&PSL.385...22T, 2014ApJ...785L..20W, 2014ApJ...792...90D, 2014ApJ...797L..25R, 2015AsBio..15..119L, 2015ApJ...806..249G, 2015NatSR...513977N, 2015ApJ...812..137H}, the simultaneous detection of large abundances of $\mathrm{O_2}$ or its photochemical byproduct $\mathrm{O_3}$ in combination with a reducing gaseous species such as $\mathrm{CH_4}$ is still considered as the most robust biosignature.
This is because since reduced and oxidizing gasses react rapidly with each other, such a detection assures a large flux of $\mathrm{O_2}$ and $\mathrm{CH_4}$ \kawashima{from the surface, therefore likely biotic in origin} \citep{1965Natur.207....9L, 1965Natur.207..568L, 1993Natur.365..715S}.
%\com{(I do not cite Lederberg 1965 since it does not seem to mention the importance of simultaneous detection of O2 and CH4 particularly.)}
Also, $\mathrm{H_2O}$, while not a biosignature, is a useful indicator of habitability.

%The atmospheric composition of the Earth has changed dramatically through its evolution.
%The origin of oxygenic photosynthesis precipitated two distinct rises at the GOE (Great Oxidation Event, sometime between 2.4 and 2.1~Ga) and NOE (Neoproterozoic Oxygenation Event, about 0.6~Ga) \citep{lyons2014rise}.
Earth's atmosphere has been very different in its history, representing a variety of possible terrestrial atmospheres \citep{2007ApJ...658..598K, 2018ApJ...854...19R}.
In addition, we expect to find atmospheric compositions far beyond what we have seen in the Earth's history or in our Solar System bodies as the detection of hot Jupiters and mini-Neptunes have already shown.
However, it is not unreasonable to search for $\mathrm{O_2}$ since the building blocks of the \kawashima{oxygenic} photosynthesis ($\mathrm{H_2O}$, $\mathrm{CO_2}$, and photons) are abundant in the Universe. 
Their widespread availability in part has made oxygenic photosynthesis the most successful biomass building strategy on the
Earth.
While $\mathrm{O_2}$ abundance in the atmospheres of habitable planets could be much less, it is likely not much more on a habitable planet with vegetation due to widespread fires if $\mathrm{O_2}$ increases above 25-35\% of the atmosphere due to widespread fires \citep{watson1978methanogenesis, scott2006diversification}.
\kawashima{Also, in Earth's history, $\mathrm{O_2}$ has not exceeded $\sim 30 - 35$\% \citep{kump2008rise, lyons2014rise}.}

The observation of flat or featureless spectra for a number of exoplanets has demonstrated the commonality of clouds and hazes \citep[e.g.,][]{2014Natur.505...69K, Sing:2016hi}.
By absorbing and scattering the light, the existence of clouds and hazes can significantly impact the spectrum of the planet \citep[e.g.,][]{2018ApJ...853....7K, 2019arXiv190210151K, Kawashima.Ikoma2019}.
%\com{(I do not want to comment on more specific results of our paper since our paper focused on haze, not clouds, which this paper focuses.)}
On Earth, the high albedo of water \kawashima{and ice} clouds compared to that of the surface can deepen molecular absorption features, while also obscuring features depending on the cloud properties (the altitude of the cloud layer and its fractional coverage) \citep[e.g.,][]{2006AsBio...6...34T, 2006AsBio...6..881T, 2007ApJ...658..598K, 2011A&A...534A..63K, 2013AsBio..13..251R}.

Previous studies have modeled the reflectance spectra of modern Earth-like planets considering the effect of clouds in the atmospheres \citep[e.g.,][]{2002AsBio...2..153D, 2006AsBio...6...34T, 2006AsBio...6..881T, 2011AsBio..11..393R, 2011A&A...534A..63K, 2013AsBio..13..251R, 2013ApJ...766..133S, 2013A&A...557A...6K, 2014ApJ...780...52S, 2015ApJ...809...57R, 2018AJ....155..200F, 2018JATIS...4c5001W}.
In addition to modern Earth-like planets, \cite{2007ApJ...658..598K} and \cite{2018ApJ...854...19R} modeled the reflectance spectra of planets similar to the Earth at earlier geological epochs orbiting around Sun-like stars, and those around F, G, K, and M stars, respectively.
While most of the above studies considered clouds with altitudes and global average coverage similar to the modern Earth, the cloud properties in other Earth-like planets is unknown and will be likely different from those of the modern Earth.
The detectability of molecular features considering such a wide range of cloud properties has not been explored in detail.

In this study, we explore the effect of water \kawashima{and ice} cloud properties, namely the altitude and its coverage, on the reflectance spectra of Earth-like planets around Sun-like stars at different geological epochs and examine the detectability of astrobiologically interesting gaseous molecules in the visible and near-infrared spectrum, namely $\mathrm{O_2}$, $\mathrm{H_2O}$, and $\mathrm{CH_4}$, with test parameters for the future mission concept, LUVOIR, using a scaled WFIRST-AFTA coronagraph noise simulator \citep{2016PASP..128b5003R}.

The rest of this paper is organized as follows. In \S\ref{sec:methods}, we describe our model. In \S\ref{sec:results}, we show the results of reflectance spectrum models of Earth-like planets at different geological epochs and systematically explore the effect of the cloud properties.
In \S\ref{sec:detectability}, we report the detectability of $\mathrm{O_2}$, $\mathrm{H_2O}$, and $\mathrm{CH_4}$ in these atmospheres using potential parameters for the future mission concept, LUVOIR. Then in \S\ref{sec:disc} and \S\ref{sec:summary}, we conclude this paper by discussing our treatment of clouds and summarizing the results.

\section{Methods} \label{sec:methods}
We simulate the reflectance spectra considering the planets with the same mass, radius, and semi-major axis as the Earth orbiting the star with the same properties as the Sun at different geological epochs. Out of four geological epochs considered in \cite{2018ApJ...854...19R}, we consider the three epochs when the Earth has had an active biosphere and oxygenic photosynthesis, 2.0~Ga, 0.8~Ga, and the present.
2.0~Ga corresponds to the time after the GOE (Great Oxidation Event) \kawashima{of $\sim 2.33$~Ga \citep[e.g.,][]{luo2016rapid}} when $\mathrm{O_2}$ started to build up in the atmosphere and 0.8~Ga corresponds to the time when multicellular life started to proliferate after the NOE (Neoproterozoic Oxidation Event).

\subsection{Reflectance Spectrum Model}
\begin{figure*}[ht!]
\plotone{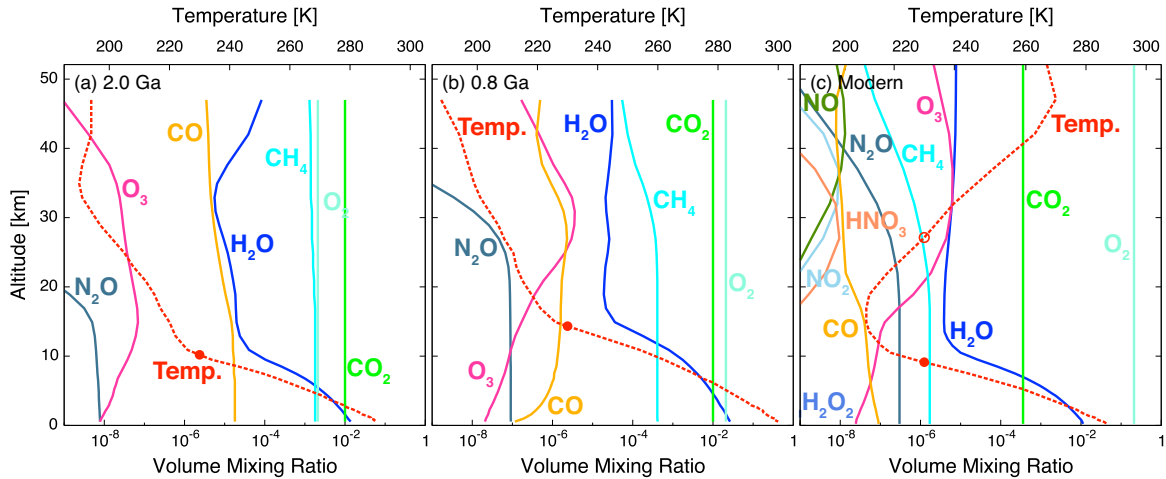}
\caption{\kawashima{Vertical profiles of temperature (red dashed line) and gaseous species (solid lines) for three different Earth-like trajectory epochs, 2.0~Ga~(a), 0.8~Ga~(b), and the modern Earth~(c). Red circles represent the 230~K threshold altitude of water and ice clouds, which are 10.3, 14.4, and 8.71~km for the cases of 2.0~Ga, 0.8~Ga, and the modern Earth, respectively. Note that most abundant species $\mathrm{N_2}$ is not shown.}\label{fig_profile}}
\end{figure*}

\begin{sloppypar}
To simulate reflectance spectra of Earth-like planets, we use a line-by-line radiative transfer model \citep{1976ApOpt..15..364T, 2009ApJ...698..519K, 2018ApJ...854...19R}.
\kawashima{We calculate the spectra with a wavenumber grid width of $0.01~\mathrm{cm^{-1}}$.}
We use the temperature-pressure profile and distribution of gaseous species of \cite{2015ApJ...806..137R} for Earth-like atmospheres at the three geological epochs as inputs to the radiative transfer model\kawashima{, which are shown in Figure~\ref{fig_profile}}.
Those results were calculated with a 1D climate model \citep{1986Sci...234.1383K, 2000JGR...10511981P, 2008AsBio...8.1127H} and a 1D photochemistry code \citep{2002AsBio...2...27P, 2005AsBio...5..706S, 2007A&A...472..665S}.
\end{sloppypar}

\kawashima{Note that the temperature and abundances for the two earlier epochs are not well-constrained and lie within an extremely broad range of possible values.}
\kawashima{We tabulate the geological constraints on the past $\mathrm{O_2}$ abundance for each geological epoch in Table~\ref{tab_abundances}.
As for $\mathrm{H_2O}$, its abundance in the atmosphere is determined by evaporation and thus surface temperature. However, considering the temperature oscillation occurred during the cooler period within the huge temporal range, it might be lower than what we assume here.}

\kawashima{For $\mathrm{CH_4}$, its past abundance in the atmosphere is not currently constrained by geological records. Photochemical model of \cite{pavlov2003methane} predicted concentration of 100-300ppm in the Proterozoic (0.75--2.3~Ga) atmosphere in order to maintain warm climate against faint early sun. Biogeochemical model of \cite{claire2006biogeochemical} derived an analytical solution of $\mathrm{CH_4}$ abundance as a function of uncertain parameters such as rate coefficient for a $\mathrm{CH_4}$ destruction by $\mathrm{O_2}$, surface biogenic flux of $\mathrm{CH_4}$, and the $\mathrm{O_2}$ abundance.
Their reference model predicted its abundance ranges from 10 to 100ppm after GOE at 2.3~Ga. 
In absence of robust geological paleosol records, we have adopted optimistic $\mathrm{CH_4}$ levels in the lowest $\mathrm{O_2}$ case. Future work will be needed to constrain {$\mathrm{CH_4}$ abundance} in Earth's history.}

\begin{deluxetable*}{lll}
\tablecaption{Geological constraints on the past $\mathrm{O_2}$ abundances \label{tab_abundances}}
\tablecolumns{3}
\tablewidth{0pt}
\tablehead{
\colhead{Epoch [Ga.]} &
\colhead{Concentration} &
\colhead{Reference}
}
\startdata
2.45-0.42 & 0.01-0.4~PAL & \cite{kump2008rise} and references therein \\
2.1-0.8 & $10^{-4}-0.1$~PAL & \cite{lyons2014rise} and references therein \\
1.8-0.8 & $< 0.001$~PAL & \cite{planavsky2014low} \\
0.42- & 0.6-1.6~PAL & \cite{kump2008rise} and references therein \\
\enddata
\tablecomments{PAL stands for the present atmospheric level.}
\end{deluxetable*}

As for clouds, we assume water (cumulus) clouds \kawashima{for temperature above 230~K and ice (cirrus) clouds for that below 230~K following \cite{2012Icar..221..603Z}.} \kawashima{We insert} continuum-absorbing/emitting layers similar to some previous works \citep{2002AsBio...2..153D, 2007ApJ...658..598K, 2013AsBio..13..251R, 2015ApJ...809...57R, 2018ApJ...854...19R}.
\kawashima{While planets with a surface ocean, an active hydrological cycle, and abundant water vapor have abundant clouds, dry habitable planets, which have been proposed to extend the habitable zone inward \citep[e.g.,][]{2005Icar..178...27A, 2011AsBio..11..443A, 2013ApJ...778..109Z, 2015ApJ...812..165K}, have fewer clouds \citep[e.g.,][]{2018JGRE..123..559K}.}
\kawashima{However, since} the cloud properties in exoplanet \kawashima{contains large uncertainty}, we \kawashima{simply} vary the altitude of the cloud layer and its coverage systematically to explore the effect of these cloud properties on reflectance spectra of Earth-like exoplanets.
%Although the clouds in the upper troposphere can be composed of ice rather than water as in the case of the present Earth, we assume the properties of water clouds for all the altitudes for simplicity.

\begin{sloppypar}
We assume surface \kawashima{compositions} following \cite{2018ApJ...854...19R}\kawashima{: The surface consists of 70\% ocean, 2\% coast, and 28\% land for all the epochs considered}. \kawashima{For 2.0~Ga and 0.8~Ga cases, the land is composed of 35\% basalt, 40\% granite, 15\% snow, and 10\% sand, while 30\% grass, 30\% trees, 9\% granite, 9\% basalt, 15\% snow, and 7\% sand for modern case.}
\kawashima{We} take reflectivity data for clouds and surface compositions from the ASTER Spectral Library\footnote{http://speclib.jpl.nasa.gov} \citep{BALDRIDGE2009711} and the USGS Spectral Library\footnote{http://speclab.cr.usgs.gov/spectral-lib.html} \citep{kokaly2017usgs}.
We adopt the average planet phase angle of $\frac{\pi}{2} $ (i.e., quadrature).
For the input stellar spectra of the Sun at each epoch, we use a solar evolution model \citep{2012ApJ...757...95C}. 
\end{sloppypar}

\subsection{LUVOIR Coronagraph Noise Simulator}
We calculate the impact of noise on the detection of spectral features considering the Earth-like planet located at 5~pc away from the Earth.
For this purpose, we use the instrument noise model from \cite{2016PASP..128b5003R} originally developed for WFIRST-AFTA. 
We have modified this noise calculator to match the potential LUVOIR values. While two plans have been proposed for the telescope diameter of LUVOIR, 15~m and 8~m, in this study, we use the value of 10~m as an example. Considering visible channel of ECLIPS instrument, we take its value for the instrument spectral resolution and coronagraph inner and outer working angle from Table~9.2 of the LUVOIR interim report.\footnote{https://asd.gsfc.nasa.gov/luvoir/} All the input values we use are listed in Table~\ref{tab_param}. Also, while the original noise model assumed the black-body for the stellar spectrum, we use a solar spectrum evolution model as the input.

Following \cite{2016PASP..128b5003R}, we explore the integration time required to detect a molecular feature by defining it as the time to achieve $\mathrm{S/N} = 5$. We define the signal as the difference between the spectra calculated with and without the specific molecular absorption, while \cite{2016PASP..128b5003R} defined it as the deviation from a flat continuum\kawashima{; we substitute the photon count rate for the case of the spectrum calculated without considering the absorption of a certain molecule for the continuum count rate in Eq.~(7) of \cite{2016PASP..128b5003R}.
The model selects a wavelength element within a specific wavelength range from a given instrument spectral resolution. We will mention the wavelength range we adopt for each molecular absorption feature in \S\ref{sec:detectability}.
Note that in order to recover molecular abundances, a measurement of the flux at the bottom of the absorption features is important.}

\begin{deluxetable*}{lll}
\tablecaption{Values of parameters used in this study \label{tab_param}}
\tablecolumns{3}
\tablewidth{0pt}
\tablehead{
\colhead{Description} &
\colhead{Value} &
\colhead{Reference}
}
\startdata
Distance to observed star-planet system & 5~pc \\
Planetary radius & 1~$R_\earth$ \\
Planet-star distance & 1~AU \\
Planet phase angle & $90^\circ$ \\
Number of exodis in exoplanetary disk & 1 \\
Coronagraph design contrast & $10^{-10}$ \\
Telescope diameter & 10~m \\
Instrument spectral resolution & 140 & LUVOIR interim report\tablenotemark{a} \\
Telescope and instrument throughput & 0.20 \\
Coronagraph inner working angle [$\lambda/D$] & 3.5 & LUVOIR interim report\tablenotemark{a} \\
Coronagraph outer working Angle [$\lambda/D$] & 64.0 & LUVOIR interim report\tablenotemark{a} \\
Width of photometric aperture [$\lambda/D$] & 1.5 \\
\enddata
\tablenotetext{a}{https://asd.gsfc.nasa.gov/luvoir/}
%\tablecomments{}
\end{deluxetable*}
\section{Results: Influence of clouds on spectra of Earth-like planets} \label{sec:results}
In this section, we systematically explore the effect of cloud properties, namely the altitude of the cloud layer (\S\ref{sec:altitude}) and its coverage (\S\ref{sec:coverage}) on reflectance spectra of an Earth-like planet. Then in \S\ref{sec:evolution}, we compare the spectrum models of the Earth-like planets at different geological epochs, focusing on the $\mathrm{O_2}$ A-band feature since $\mathrm{O_2}$ has long been considered as a key target molecule for future missions.

\subsection{Altitude of Cloud Layer} \label{sec:altitude}
\begin{figure}[ht!]
\includegraphics[width = 0.48\textwidth]{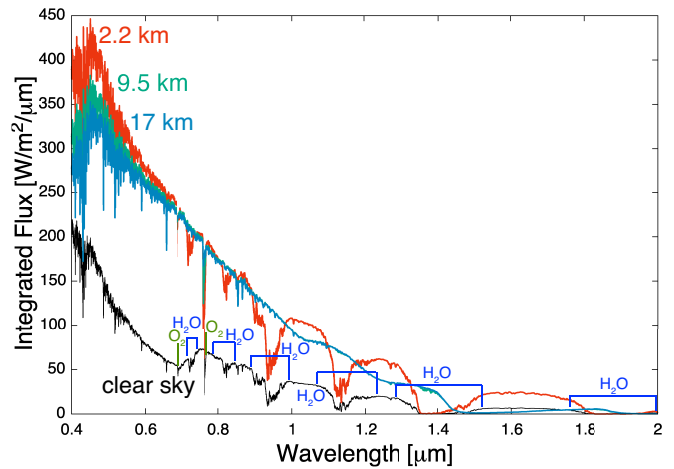}
\caption{Reflective Spectra for a modern Earth-like atmosphere with 100\% cloud coverage at three different altitudes, 
17~km~(blue line), 9.5~km~(green line), and 2.2~km~(red line). \kawashima{Water clouds are assumed for 2.2~km case, while ice clouds for 17 and 9.5~km cases.} A clear sky atmosphere is plotted in black line for reference. \kawashima{Note that the spectral models are smoothed for clarity by averaging over the wavenumber range of $20.1~\mathrm{cm}^{-1}$ at each outputted wavenumber point with a grid of $0.1~\mathrm{cm}^{-1}$. We use the same smoothing method for the results of spectrum models hereafter.}\label{fig_altitude}}
\end{figure}

Figure~\ref{fig_altitude} shows spectral models for an atmosphere with 100\% cloud coverage at three different altitudes, 17~km~(blue line), 9.5~km~(green line), and 2.2~km~(red line). \kawashima{Water clouds are assumed for 2.2~km case, while ice clouds for 17 and 9.5~km cases.}
A clear sky atmosphere is also plotted (black) for reference.
\kawashima{One finds that in the spectrum of a clear sky atmosphere,} most of the molecular absorption features come from $\mathrm{H_2O}$, which are located at 0.71-0.74, 0.80-0.84, 0.90-0.98, 1.1-1.2, 1.3-1.5, and 1.8-2.0~$\mu$m, while the distinct $\mathrm{O_2}$ A-band feature exists at 0.76~$\mu$m \kawashima{along with smaller $\mathrm{O_2}$ B-band feature at 0.69~$\mu$m}. 

Clouds increase the flux because of their high albedo.
\kawashima{At relatively short wavelengths ($\lesssim 0.9$~$\mu$m), where the atmosphere is relatively optically thick and the optical properties of water and ice clouds are almost similar,} the lower the altitude of the cloud layer is, the larger the overall (continuum) flux becomes. This behavior is due to the increased Rayleigh scattering of molecules above the cloud layer in the lower atmosphere.
%\com{(I do not think that the high albedo of the clouds compared to the relatively low reflectance of the surface is also the reason for higher flux for lower cloud layers because the same cloud coverage is assumed for the cloud layers at all the three altitudes. It is of course the reason for higher flux for cloudy atmosphere compared to clear atmosphere.)}
For the lower altitude clouds, the absorption feature is deeper, and the flux is lower in the core of the line. This is because there is a larger column-integrated concentration of the species above the cloud layer \citep[see also][]{2006AsBio...6...34T, 2006AsBio...6..881T, 2011A&A...534A..63K}.
%This is because the difference of the altitude important for the spectrum (namely, where optical depth, $\tau \sim 1$) between wavelengths at the peak of the absorption features and continuum becomes larger for lower cloud layer.

\kawashima{In contrast, at relatively long wavelengths ($\gtrsim 0.9$~$\mu$m), where the atmosphere is optically thinner, the features are created mostly by clouds, while the molecular absorption also contribute for the lower altitude cloud case of 2.2~km. Note that water clouds have absorption at the similar wavelength region as gaseous water. For the higher altitude ice cloud cases of 17 and 9.5~km, due to the negligible column-integrated concentration of the species above the cloud layer for the both cases, the spectra are similar and completely characterized by less reflective optical properties of ice clouds.}

\begin{figure*}[ht!]
\plottwo{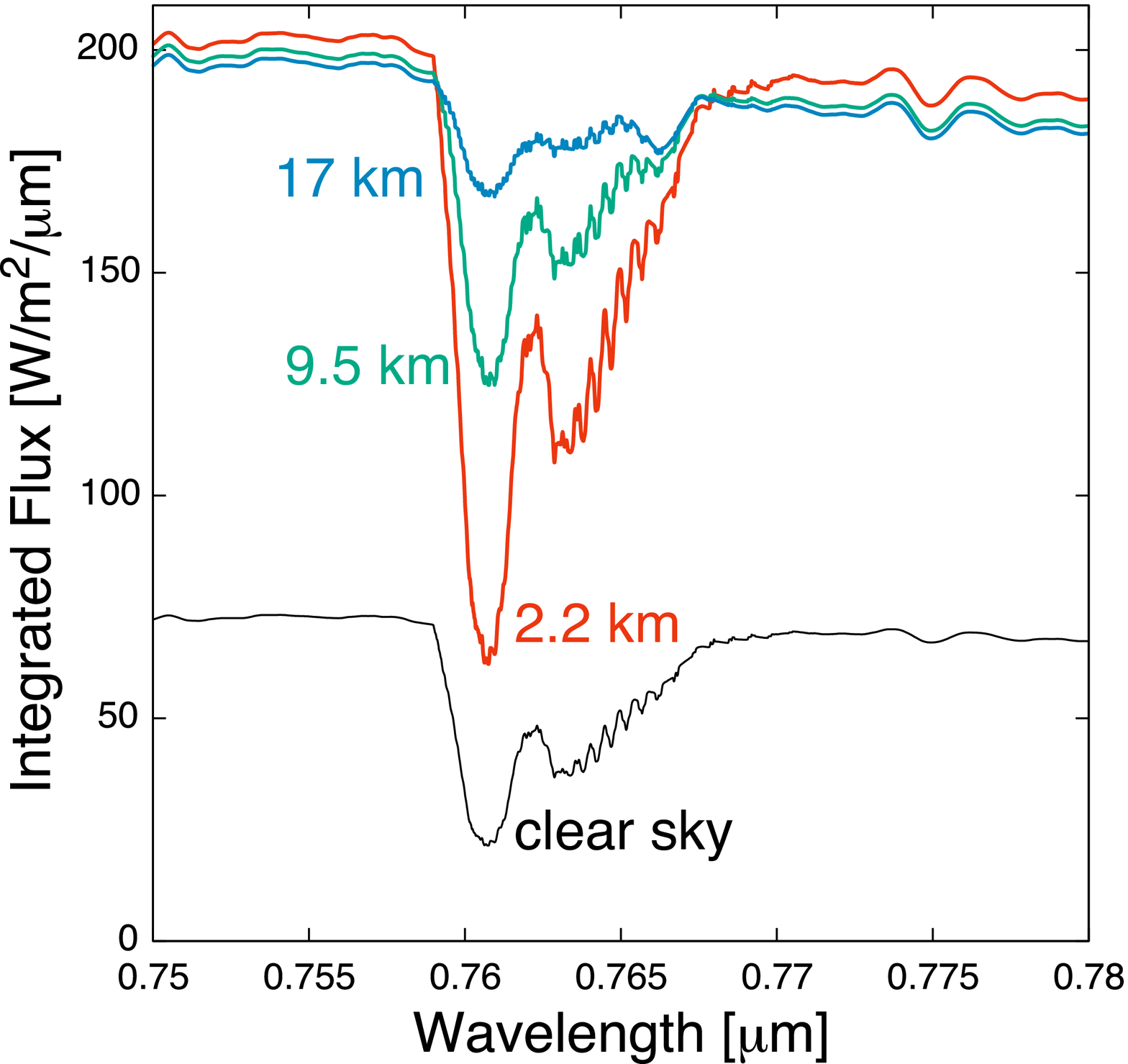}{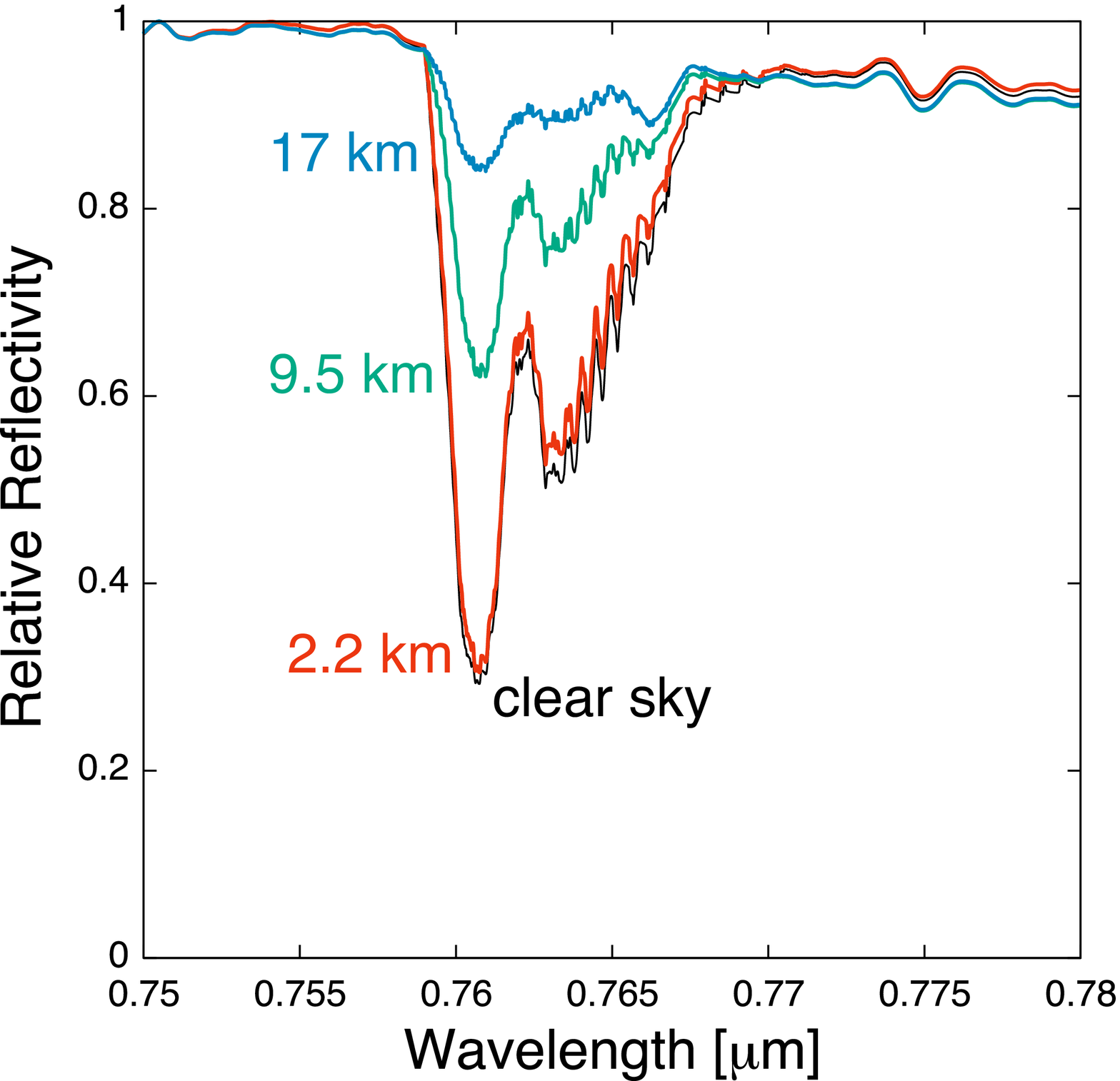}
\caption{Same as Fig.~\ref{fig_altitude} with only the $\mathrm{O_2}$ A-band feature shown in integrated flux (left) and relative reflectivity (right).
The relative reflectivity is calculated by normalizing with the maximum flux between the wavelength range of 0.75-0.78~$\mu$m.\label{fig_O2_altitude}}
\end{figure*}

The left panel of Figure~\ref{fig_O2_altitude} is the zoomed in view of Fig.~\ref{fig_altitude} around the $\mathrm{O_2}$ A-band feature.
\kawashima{Note that the difference of the reflectivity of water and ice clouds is little in this wavelength region.}
The flux at the peak of the absorption feature is smaller for the lower cloud layer, while that at the continuum is larger as noted above.
We show relative reflectivity in the right panel of Figure~\ref{fig_O2_altitude} calculated by normalizing the flux with the maximum flux between the wavelength range of 0.75-0.78~$\mu$m. For the lower altitude clouds, the relative reflectivity of the feature becomes deeper due to the larger absorption at the core of the feature and increased Rayleigh scattering at the continuum.

\subsection{Cloud Coverage} \label{sec:coverage}
\begin{figure*}[ht!]
\plotone{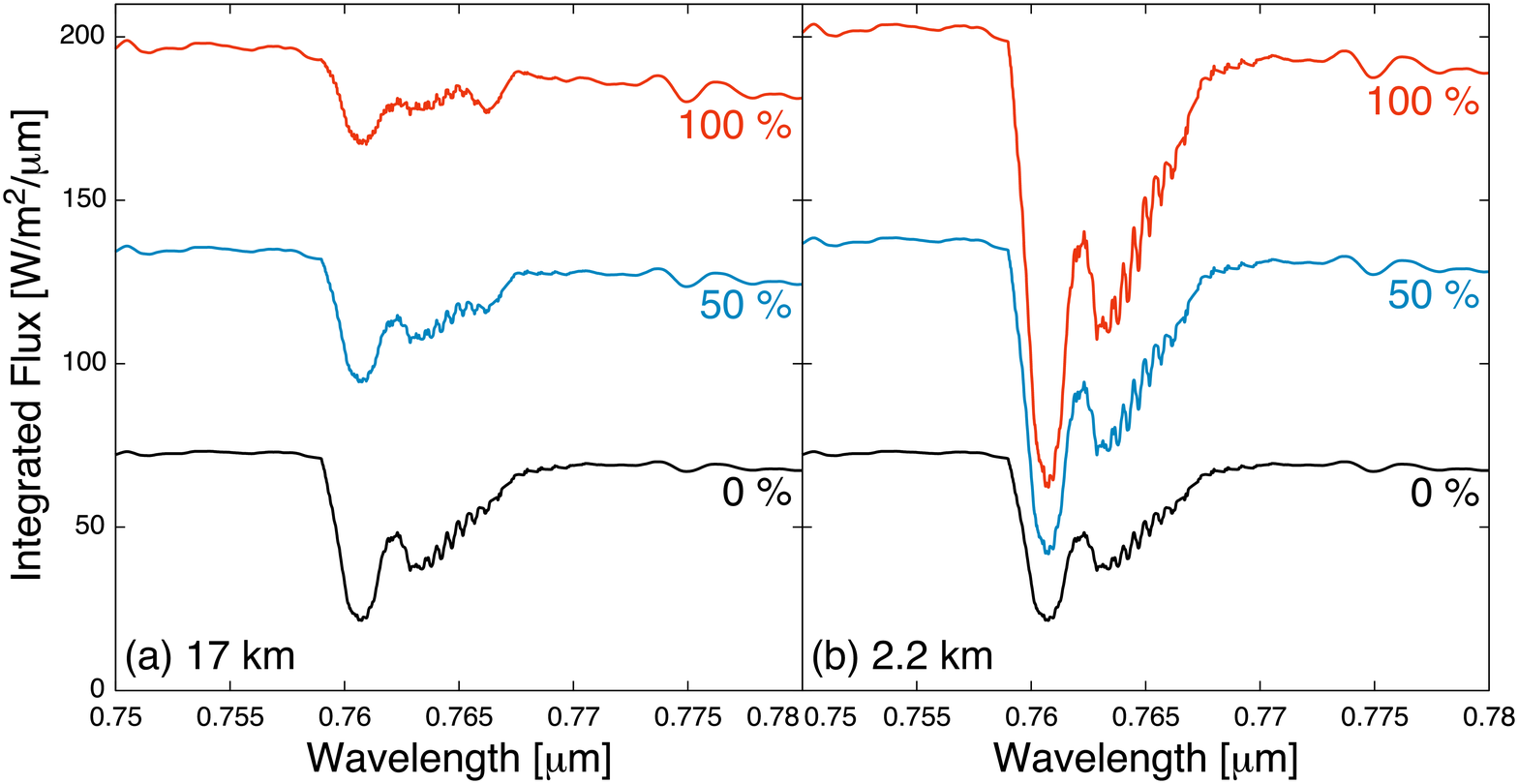}
\plotone{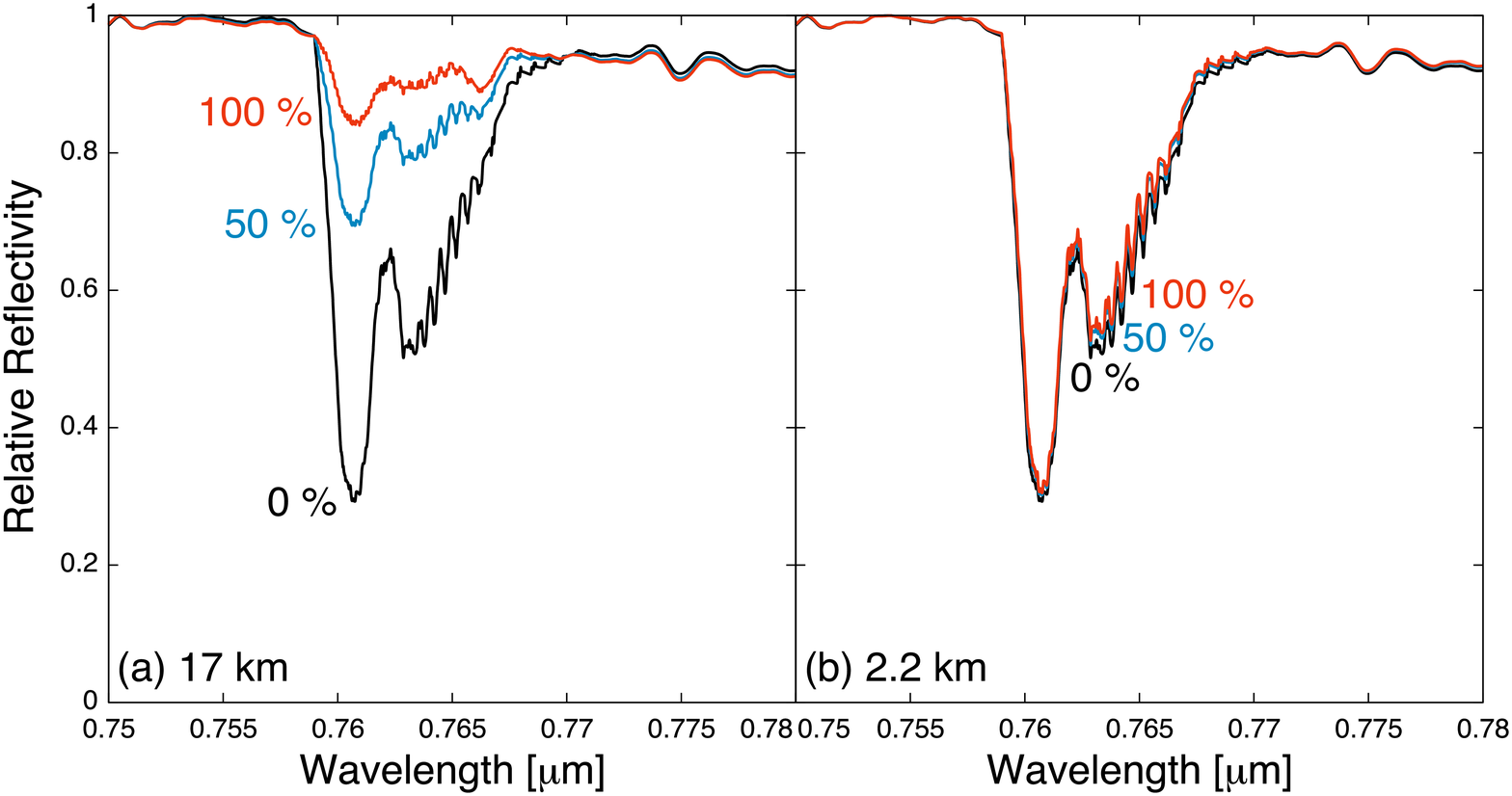}
\caption{The $\mathrm{O_2}$ A-band with \kawashima{ice} cloud layers of 17 km (left two panels) and 2.2 km altitude \kawashima{water} cloud layers (right two panels) with 0\%~(black), 50\%~(blue), and 100\%~(red) cloud coverage plotted with integrated flux (top two panels) and relative absorption (bottom two panels).\label{fig_O2_coverage}}
\end{figure*}

Next, we examine the dependence of the fractional cloud coverage on the spectra.
The top left panel~(a) of Figure~\ref{fig_O2_coverage} shows the Earth-like spectra with \kawashima{ice} cloud layers of 17~km altitude, while the top right one~(b) shows those with 2.2~km \kawashima{water} cloud layers, for 0\%~(black), 50\%~(blue), and 100\%~(red) cloud coverage.
\kawashima{Again, the difference of the reflectivity of water and ice clouds is little in this wavelength region.}
For the 17~km cloud case~(a), the flux at the depth of the absorption feature varies more with cloud coverage than compared to the 2.2~km case~(b) because the flux at the core of the feature is determined by the amount of the absorption, namely column-integrated O$_2$ concentration of the species above the cloud layer. The continuum increases with increasing cloud coverage due to the higher albedo of water clouds compared to the surface reflectivity.

The bottom two panels of Fig.~\ref{fig_O2_coverage}~(c, d) are the same as the top two panels of Fig.~\ref{fig_O2_coverage}~(a, b), but with relative reflectivity. It can be seen that for the 17~km case~(c), the relative absorption varies greatly with the cloud coverage and is deeper for the lower cloud coverage due to blocking more of the atmosphere below the cloud layer.
While for the 2.2~km case~(d), the relative reflectivity hardly varies with the cloud coverage although it is slightly shallower for the higher cloud coverage.

Our results for the modern Earth case confirm previous findings by \cite{2006AsBio...6...34T, 2006AsBio...6..881T, 2007ApJ...658..598K, 2011A&A...534A..63K}. We will now consider the case of earlier geological epochs in \S\ref{sec:evolution} and calculate the detectability of these features with a LUVOIR sized telescope in \S\ref{sec:detectability}.

\subsection{Evolution of the Planet} \label{sec:evolution}
In this section, we explore the spectra of an Earth-like planet at different levels of oxygen and geological epochs.
The abundance of $\mathrm{O_2}$ in the Earth's atmosphere has varied over time but broadly rose after two oxygenation events known as the Great Oxygenation Event (GOE) and the Neoproterozoic Oxygenation Event (NOE) \citep{lyons2014rise}. \kawashima{We adopt} concentrations of 0.01~PAL, 0.1~PAL, and 1.0~PAL for 2.0~Ga, 0.8~Ga, and the present, respectively, where PAL stands for the present atmospheric level. Note that oxygen levels during the Proterozoic are debated and estimates range from \kawashima{$< 0.001$~PAL} to 0.4~PAL \citep{2005AREPS..33....1C, kump2008rise, planavsky2014low} \kawashima{as listed in Table~\ref{tab_abundances}}.
%We have chosen the more optimistic oxygen level at 2.0~Ga for this study.
To explore the effect of $\mathrm{O_2}$ abundance on the spectra in detail, we also consider the case of 0.5~PAL $\mathrm{O_2}$ as a middle value. We calculate the spectrum model of the 0.5~PAL case using the same inputs to the radiative transfer model as modern Earth except for $\mathrm{O_2}$ abundance. Note this treatment is valid as long as one compares the spectrum models only around the wavelength range of $\mathrm{O_2}$ absorption features.

\begin{figure*}[ht!]
\plottwo{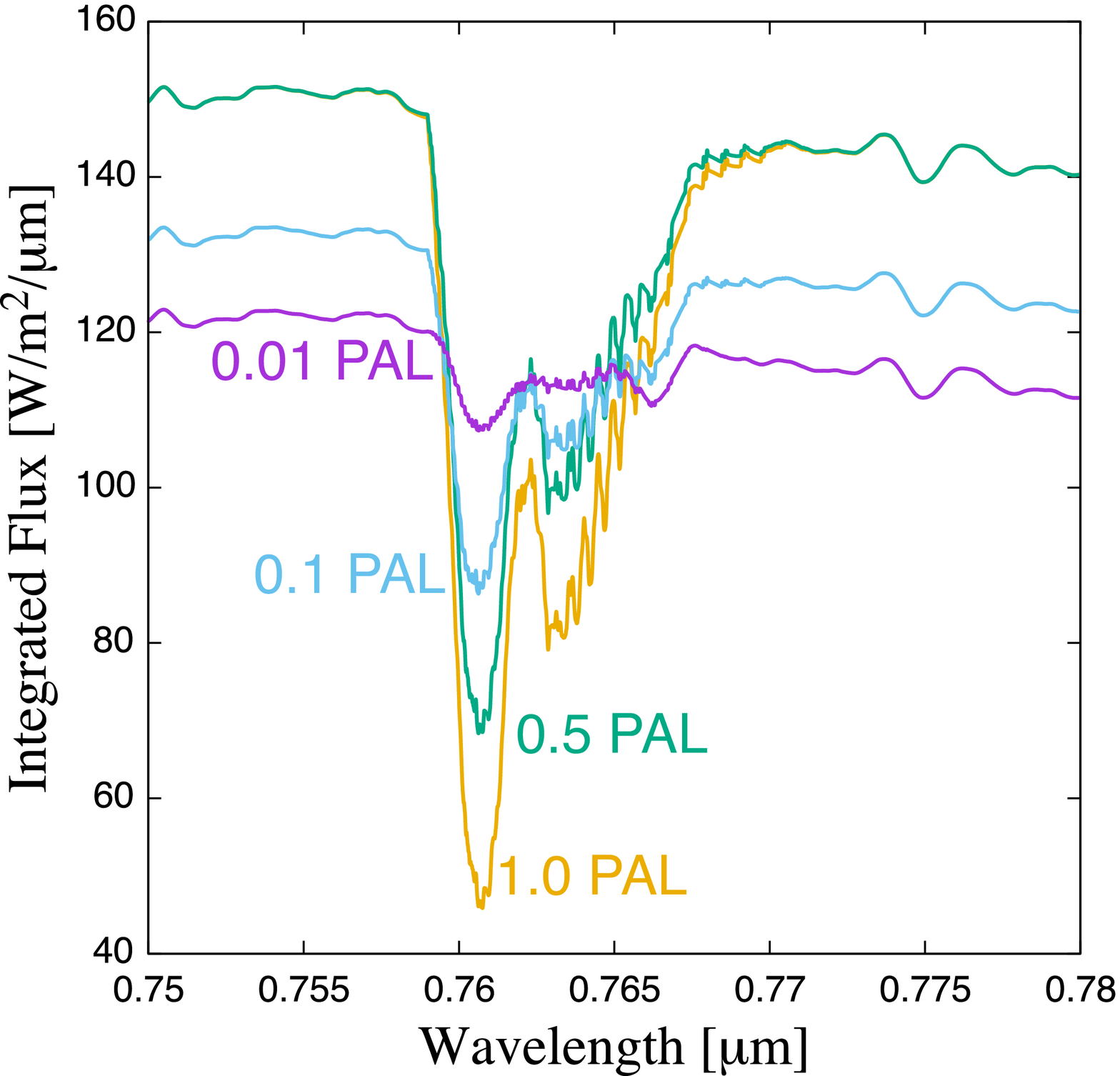}{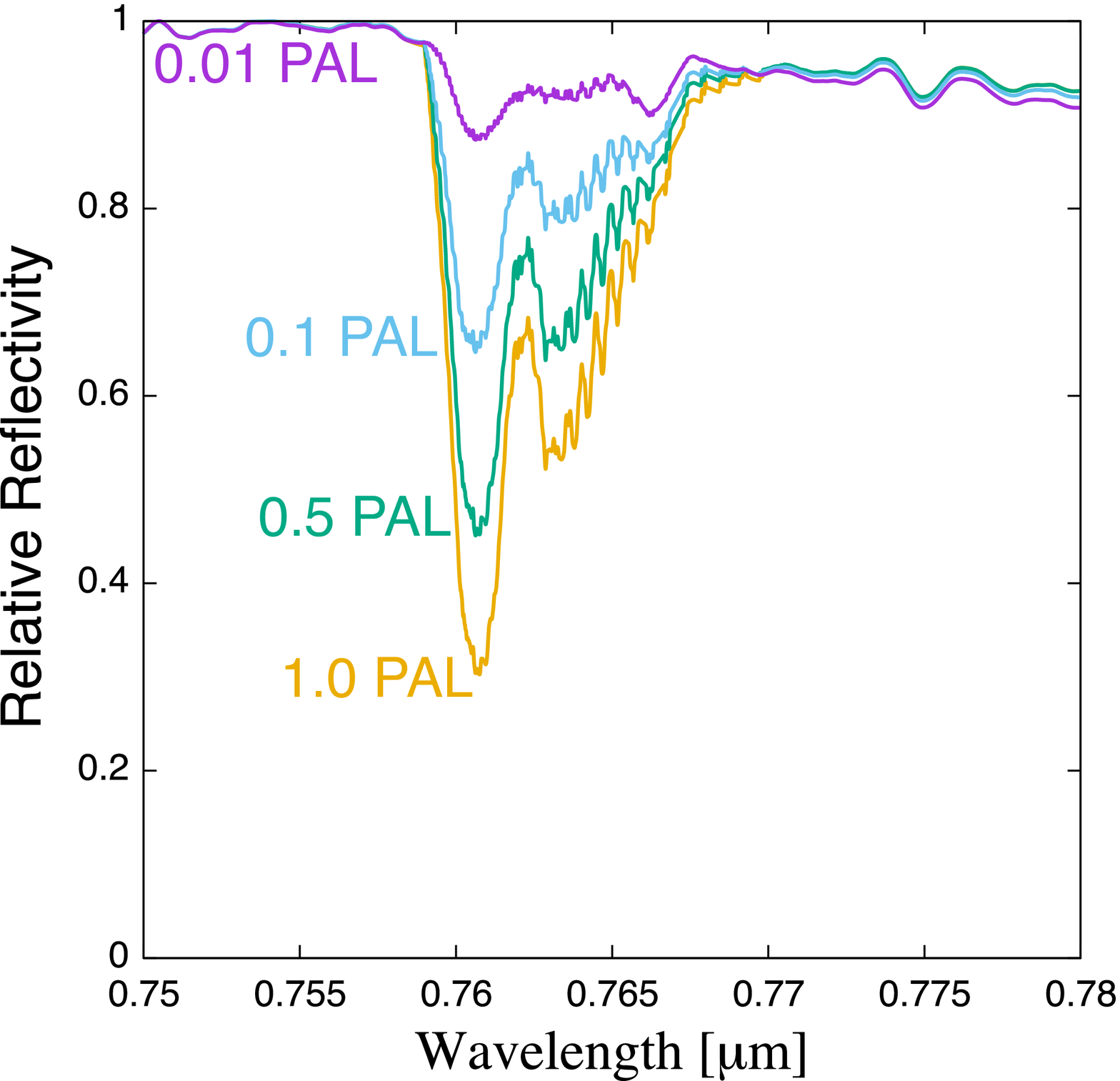}
\caption{Integrated Flux (left) and relative reflectivity (right) spectral models for four different $\mathrm{O_2}$ abundances, 0.01~PAL~(2.0~Ga, purple), 0.1~PAL~(0.8~Ga, light blue), 0.5~PAL~(green), and 1.0~PAL~(modern, orange) assuming a 2.2~km \kawashima{water} cloud layer altitude with 60\% cloud coverage.\label{fig_O2_abundance}}
\end{figure*}

Figure~\ref{fig_O2_abundance} shows the spectrum for four different $\mathrm{O_2}$ abundance models, 0.01~PAL~(2.0~Ga, purple), 0.1~PAL~(0.8~Ga, light blue), 0.5~PAL~(green), and 1.0~PAL~(0.0~Ga, orange) assuming 60\% cloud coverage with a 2.2~km \kawashima{water} cloud layer. As expected, the absorption feature is deeper for larger $\mathrm{O_2}$ abundance.
%Note the depth of the feature, namely the decrease in the relative reflectivity around the feature, does not vary linearly with the amount of $\mathrm{O_2}$. %\com{(It does not vary linearly since the depths are about 0.1, 0.3, 0.5, and 0.7 for 0.01 PAL, 0.1 PAL, 0.5 PAL, and 1.0 PAL cases, respectively.)}

\section{Results: Detectability of $\mathrm{O_2}$, $\mathrm{H_2O}$, and $\mathrm{CH_4}$ with LUVOIR} \label{sec:detectability}
In this section, we explore the detectability of the features of astrobiologically important gaseous molecules in the visible and near-infrared region of the Earth-like spectrum, namely $\mathrm{O_2}$, $\mathrm{H_2O}$, and $\mathrm{CH_4}$ with proposed space telescope LUVOIR.

\begin{figure}[ht!]
\includegraphics[width = 0.48\textwidth]{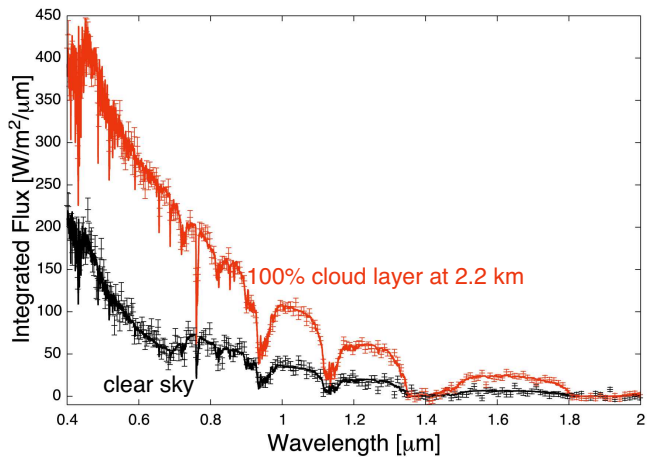}
\caption{Modern Earth-like spectra of a clear sky atmosphere~(black) and the same atmosphere with a 100\% \kawashima{water} cloud coverage layer at 2.2~km~(red) along with 1$\sigma$ observational errors for 10-hour observation with a LUVOIR-sized telescope calculated with the noise model. The assumed distance to the planetary system is 5~pc. \label{fig_cloud_noise}}
\end{figure}

Figure~\ref{fig_cloud_noise} shows the modern Earth-like spectra of a clear sky atmosphere~(black) and the same atmosphere with a 100\% \kawashima{water} cloud coverage layer at 2.2~km~(red) along with 1$\sigma$ observational errors for 10-hour observation with a LUVOIR-sized telescope calculated with the noise model. The assumed distance to the planetary system is 5~pc.
Note the negative flux means that the measurement is consistent with zero flux since a Sun-like star has low flux in the NIR.

\subsection{$\mathrm{O_2}$ Feature \label{detec_o2}}
\begin{figure*}[ht!]
\plotone{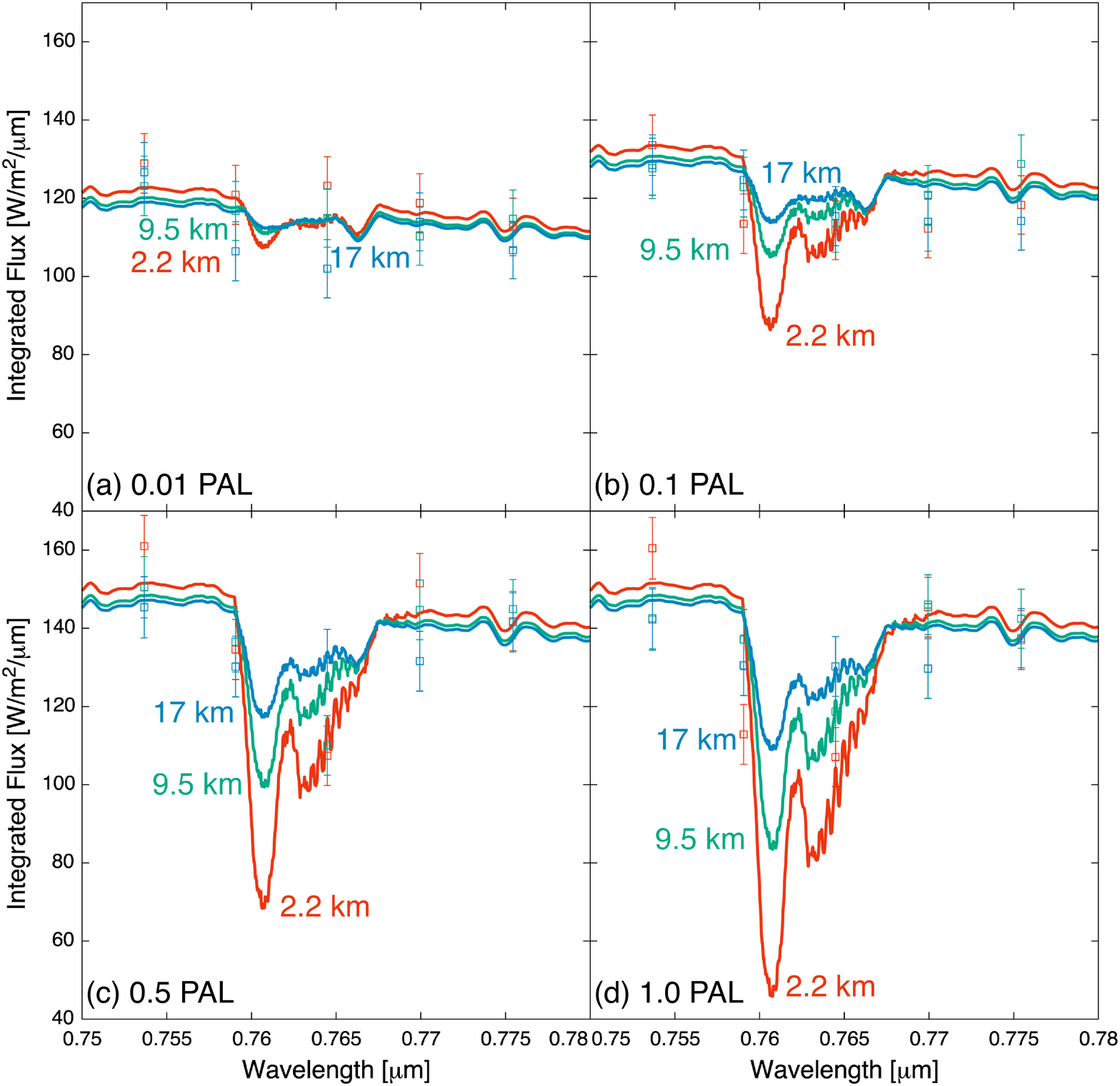}
\caption{$\mathrm{O_2}$ A-band feature for the Earth-like atmosphere with 60\% cloud coverage at different altitudes, 17~km~(blue), 9.5~km~(green), and 2.2~km~(red) with 1$\sigma$ observational errors for 10-hour observation calculated with the noise model for four different $\mathrm{O_2}$ abundances, 0.01~PAL~(a), 0.1~PAL~(b), 0.5~PAL~(c), 1.0~PAL~(d). \kawashima{Water clouds are assumed for the cases of 9.5 and 2.2~km cloud layers of 0.01 and 0.1~PAL $\mathrm{O_2}$ abundances and 2.2~km cloud layers of 0.5 and 1.0~PAL $\mathrm{O_2}$ abundances, while ice clouds are assumed for the other cases.} \label{fig_O2_altitude-noise}}
\end{figure*}

Figure~\ref{fig_O2_altitude-noise} shows spectrum models for the Earth-like atmosphere with 60\% cloud coverage at different altitudes, 17~km~(blue), 9.5~km~(green), and 2.2~km~(red) around the $\mathrm{O_2}$ A-band feature with 1$\sigma$ observational errors for 10-hour observation calculated with the noise model for four different $\mathrm{O_2}$ abundances, 0.01~PAL~(a), 0.1~PAL~(b), 0.5~PAL~(c), 1.0~PAL~(d).
\kawashima{Water clouds are assumed for the cases of 9.5 and 2.2~km cloud layers of 0.01 and 0.1~PAL $\mathrm{O_2}$ abundances and 2.2~km cloud layers of 0.5 and 1.0~PAL $\mathrm{O_2}$ abundances, while ice clouds are assumed for the other cases.} {The assumed distance to the planetary system is 5~pc.}

\kawashima{Again, note that the difference of the reflectivity of water and ice clouds in this wavelength region is minimal.}
\kawashima{We also note that we present the results on the grids we run the simulations and the stark contour lines come from our low-resolution grids.}
%\com{(It is usually given as 1$\sigma$.)}

\kawashima{We find} that the observational 1$\sigma$ error bars are much larger than the $\mathrm{O_2}$ absorption feature depth for the 0.01~PAL $\mathrm{O_2}$ concentration case~(a), but comparable or smaller for larger $\mathrm{O_2}$ concentration cases of 0.1~PAL~(b), 0.5~PAL~(c), and 1.0~PAL~(d), especially for the cases of cloud layers at the lower altitudes. The integration time required to detect the $\mathrm{O_2}$ A-band feature with the proposed LUVOIR telescope with $\mathrm{S/N} = 5$ for the 2.2~km altitude cloud layer and 0.5~PAL $\mathrm{O_2}$ concentration case~(red line in Fig.~\ref{fig_O2_altitude-noise}c) is 9.4~hour, almost the same as the assumed observation time. \kawashima{Here, we assume the wavelength region of the feature is 0.759-0.769~$\mathrm{\mu}$m.}
The detection time for each case in Fig.~\ref{fig_O2_altitude-noise} is tabulated in Table~\ref{tab_time}. 
%\com{(I need to specify the wavelength range because I calculated the detection time integrating the flux difference between this specific wavelength range and since this wavelength range affects the detection time significantly, I need to state it with triple digits. I have chosen the wavelength range which gives the shortest detection time.)}
%Note that it would take 17.5~hour to detect an ozone feature at 0.6~$\mu$m with $\mathrm{S/N} = 5$ in the atmosphere of an modern Earth at 10~pc with a 6.5~m space-based telescope like JWST, which we derive from the results of \cite{2009ApJ...698..519K}. \kawashima{}
%\com{(As you pointed out, ozone feature at 0.6~$\mu$m \cite{2009ApJ...698..519K} explored is at the edge of the instrument of JWST, so it won't perform well. I found a paper (Barstow & Irwin 2016) reporting that if modern Earth levels of ozone is present in the atmospheres of 12~pc away TRAPPIST-1 planets, it would take 60 transits for the innermost planet 1b, and 30 transits for 1c and 1d for the detection of its feature at 9~$\mu$m. But I excluded the sentences regarding the detection of ozone since it might not be so relevant.)}

\begin{deluxetable}{rrrr}[ht!]
\tablecaption{Integration time [hour] required to detect $\mathrm{O_2}$ A-band feature of 0.759-0.769~$\mathrm{\mu}$m in the atmosphere of Earth-like planet located at 5~pc away with the proposed LUVOIR telescope with $\mathrm{S/N} = 5$ for the cases of three different cloud layer altitudes, 17, 9.5, and 2.2~km and four different $\mathrm{O_2}$ abundances, 0.01, 0.1, 0.5, 1.0~PAL. The assumed cloud coverage is 60\%.\label{tab_time}}
\tablecolumns{4}
\tablewidth{0pt}
\tablehead{
\colhead{} & \multicolumn{3}{c}{Altitude of cloud layer} \\
\colhead{$\mathrm{O_2}$ abundance} &
\colhead{17~km} &
\colhead{9.5~km} & 
\colhead{2.2~km}
}
\startdata
0.01 PAL & 4800 & 1800 & 470 \\
0.1 PAL & 380 & 130 & 33 \\
0.5 PAL & 85 & 30 & 9.4 \\
1.0 PAL & 47 & 16 & 5.2 \\
\enddata
%\tablecomments{}
\end{deluxetable}

\begin{figure*}[ht!]
\gridline{\fig{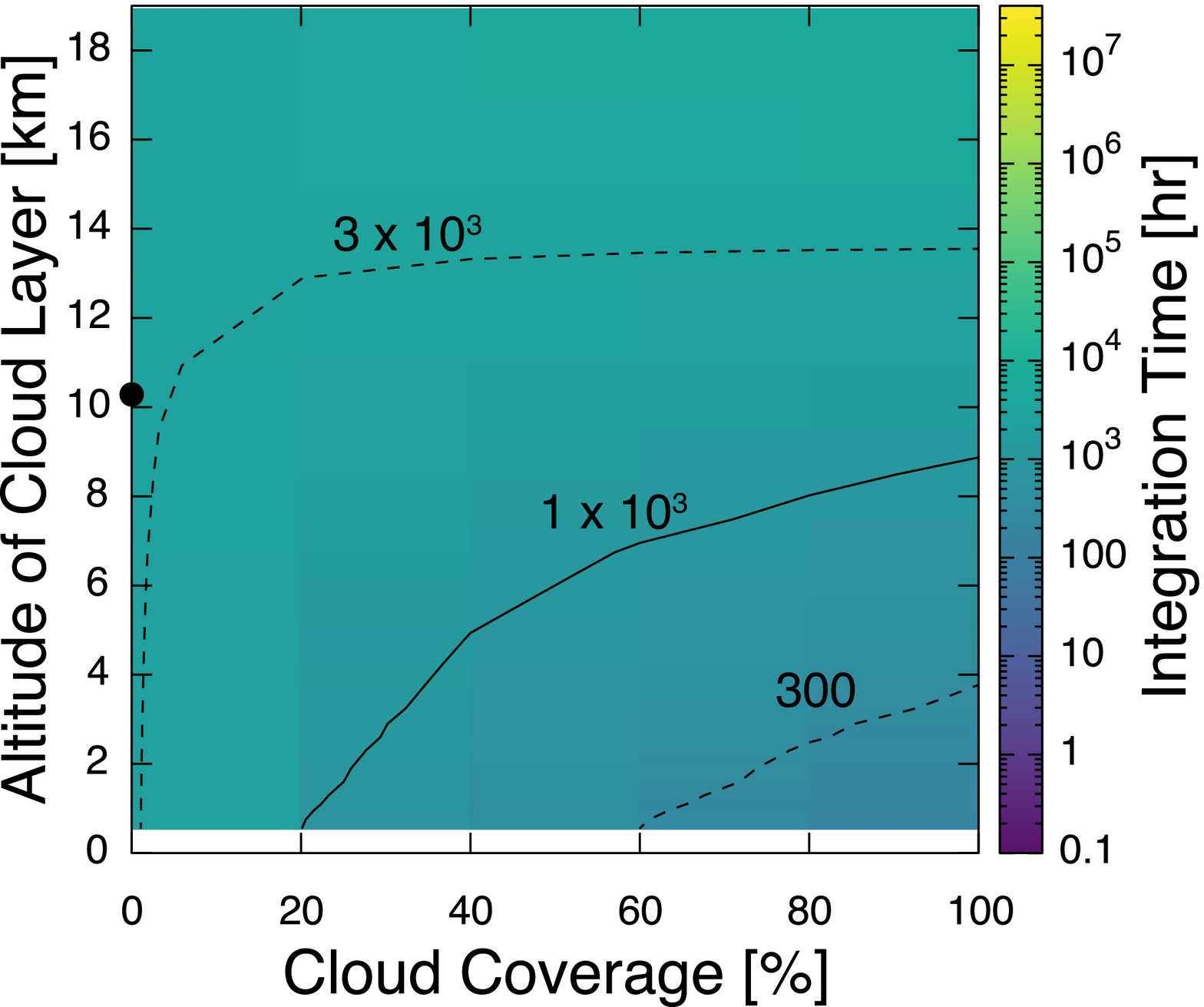}{0.4\textwidth}{(a)~0.01~PAL (2.0~Ga)}
          \fig{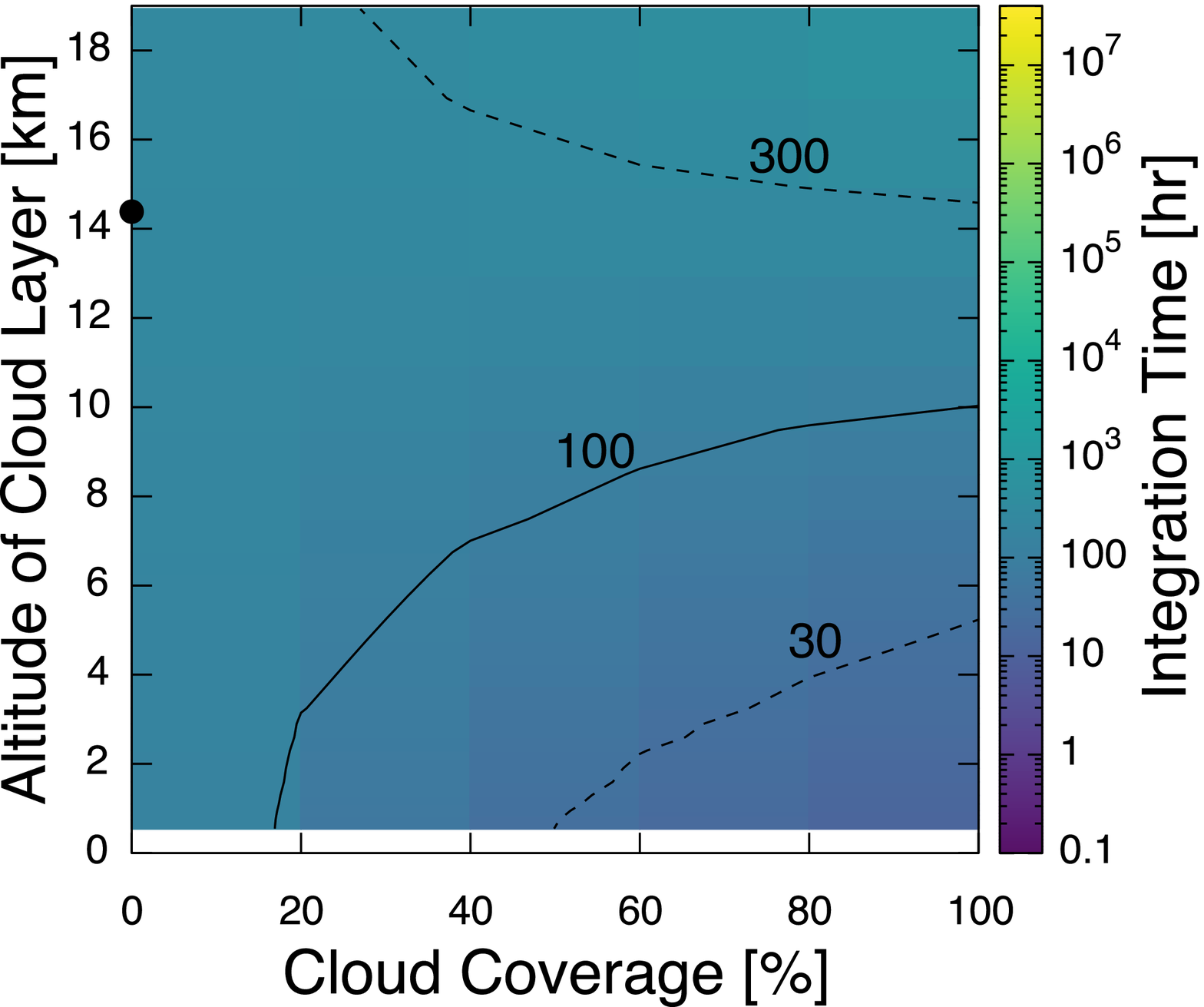}{0.4\textwidth}{(b)~0.1~PAL (0.8~Ga)}
          }
\gridline{\fig{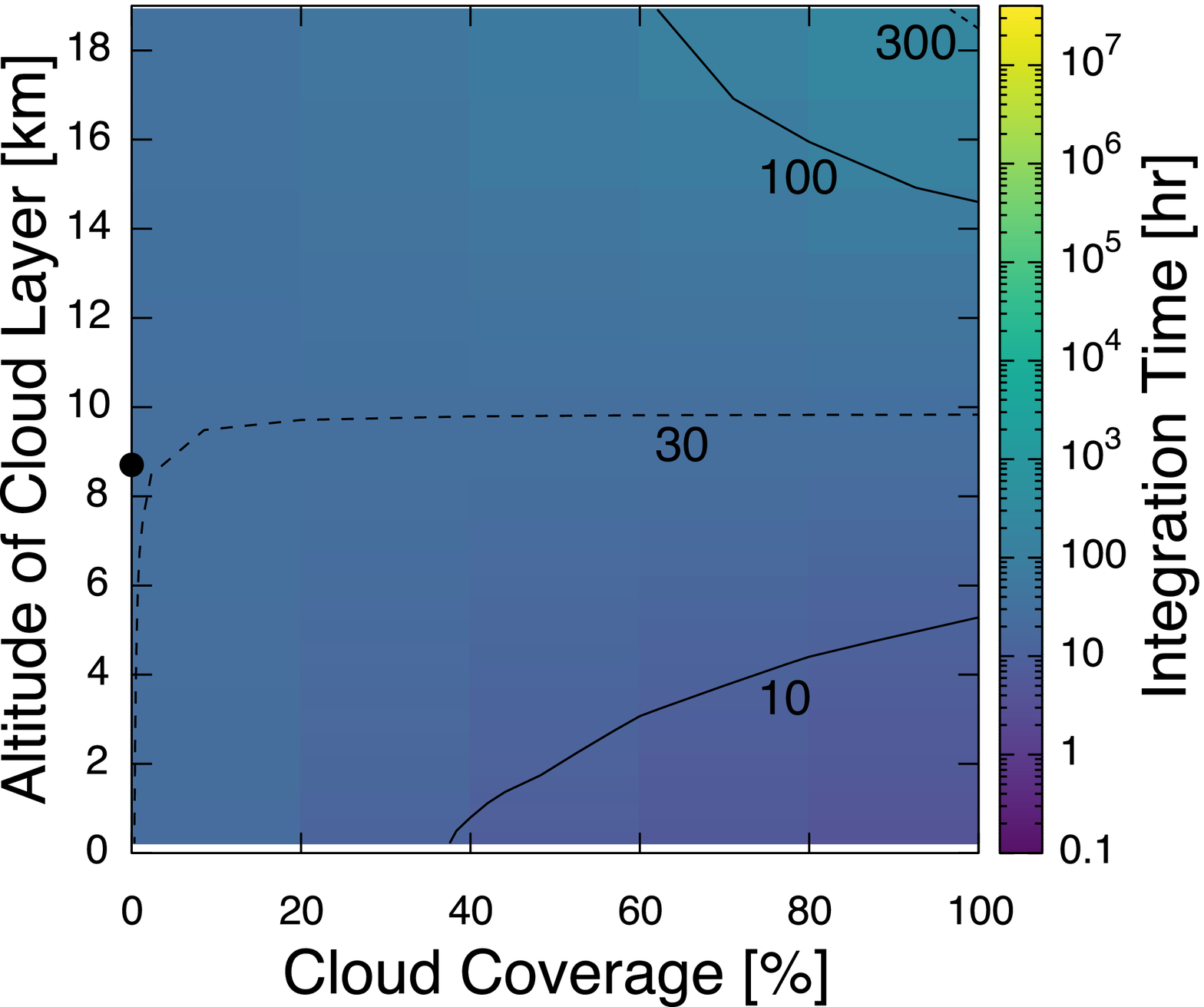}{0.4\textwidth}{(c)~0.5~PAL}
          \fig{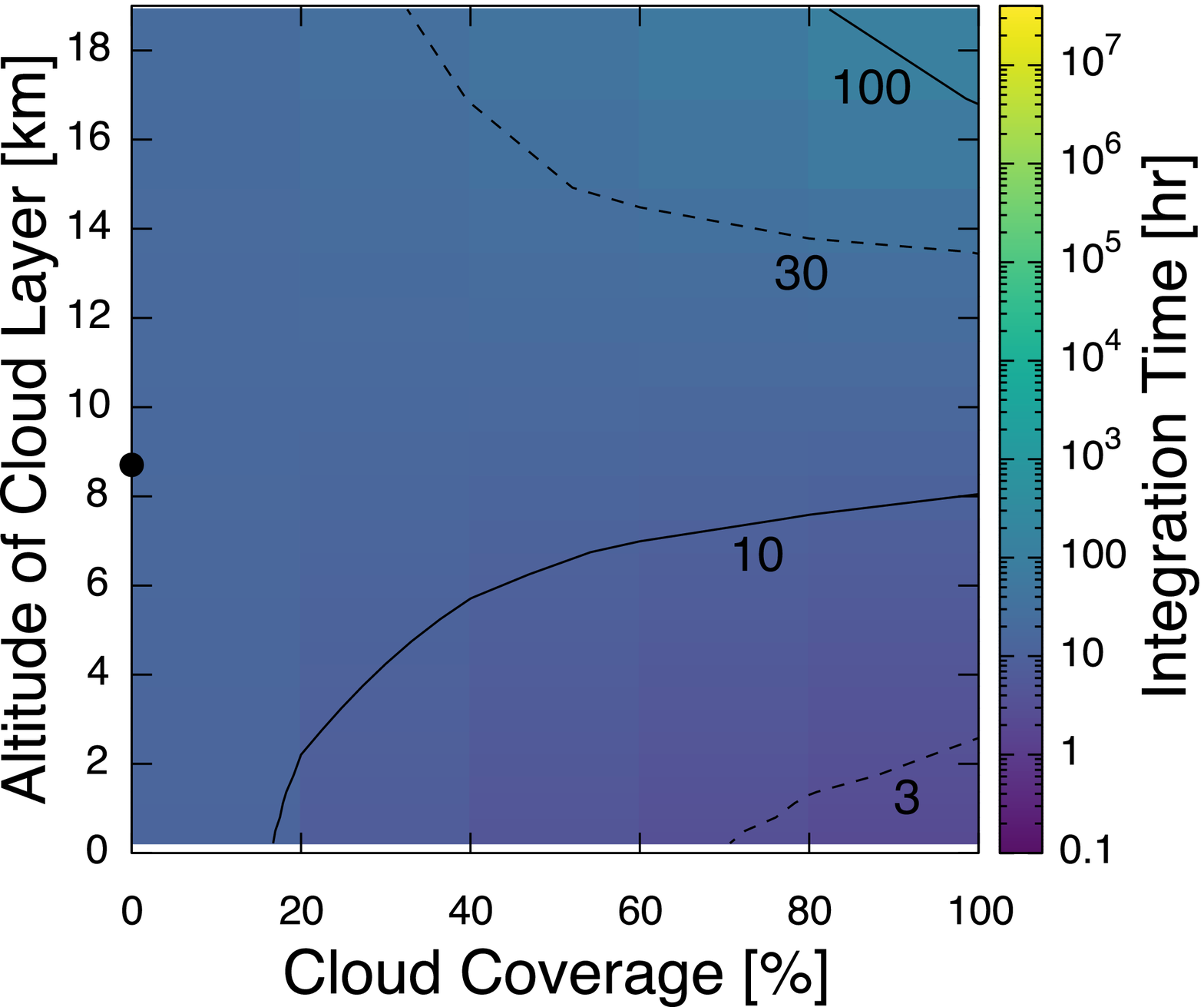}{0.4\textwidth}{(d)~1.0~PAL (modern)}
          }
\caption{Intensity plots for the integration time required to detect the $\mathrm{O_2}$ A-band feature of 0.759-0.769~$\mathrm{\mu}$m with $\mathrm{S/N} = 5$ are shown for four different $\mathrm{O_2}$ abundances, 0.01~PAL~(2.0~Ga)~(a), 0.1~PAL~(0.8~Ga)~(b), 0.5~PAL~(c), and 1.0~PAL~(modern)~(d) with varying cloud altitude and coverage. Contour lines for the integration time are also plotted in black solid and dashed lines. \kawashima{Also, the black filled circles on the vertical axes represent the altitude at 230~K, above which clouds are assumed as ice ones.} \label{fig_detectability}}
\end{figure*}

Figure~\ref{fig_detectability} shows an intensity plot for the integration time required to detect the $\mathrm{O_2}$ A-band feature with $\mathrm{S/N} = 5$ for four different $\mathrm{O_2}$ abundances, 0.01~PAL~(a), 0.1~PAL~(b), 0.5~PAL~(c), and 1.0~PAL~(d) with varying cloud altitude and coverage. 
\kawashima{The assumed wavelength region for the feature is 0.759-0.769~$\mathrm{\mu}$m.}
For high altitude ($\gtrsim 10$~km for the modern case) clouds, a lower cloud coverage makes the feature deeper and the detection easier despite the smaller continuum flux for a lower cloud coverage (see Fig.~\ref{fig_O2_coverage}a,c).
For low altitude ($\lesssim 10$~km for the modern case) clouds, a higher cloud coverage increases the flux at the continuum, while almost the same relative depth of the feature regardless of the cloud coverage, and thus makes the detection easier for a higher cloud coverage (see Fig.~\ref{fig_O2_coverage}b, d). 

As seen in Fig.~\ref{fig_detectability}, for the proposed LUVOIR telescope the integration time needed to detect the $\mathrm{O_2}$ A-band feature for an Earth-like atmosphere with $\mathrm{O_2}$ abundance of 0.01~PAL~(a) will take typically more than 1000~hours for the majority of the cloud parameters. The best case scenario would be for a widespread low layer cloud, which would then make the feature detectable with 100~hours. 
For 0.1~PAL, 0.5~PAL, and 1.0~PAL O$_2$ cases, for the majority of the cloud parameter space the detection will take approximately 100, 30, and 10~hours, respectively (see Fig~\ref{fig_detectability}). For the cloud parameter end cases, the minimum and maximum detection times are 10 to 600~hours for the 0.1~PAL case, 3 to 300~hours for the 0.5~PAL case, and 2 to 200~hours for the 1~PAL O$_2$ case. Especially, for the atmospheres with 0.5 and 1.0~PAL $\mathrm{O_2}$ abundances, the feature will be detectable with integration time $\lesssim 10$~hours as long as there are lower altitude ($\lesssim 8$~km) clouds with a global coverage of $\gtrsim 20\%$.
Note that modern Earth has a global cloud coverage of $\sim$50-60\% \citep{2006AsBio...6...34T, 2011AsBio..11..393R}.
%\kawashima{(I would like to cite EPOXI paper here as you said. In your 2013 paper, you wrote "Atmospheric models found the best match to be for a 50\% cloud coverage with 1.5 km and 8.5 km cloud layer, respectively (Robinson et al., 2011).", but I cannot find this statement in Robinson et al. 2011. Could you tell me where it is written?)}

\subsection{$\mathrm{H_2O}$ Feature}
Among the several $\mathrm{H_2O}$ features at 0.71-0.74, 0.80-0.84, 0.90-0.98, 1.1-1.2, 1.3-1.5, and 1.7-2.0~$\mu$m, we explore the detectability of the strongest feature at 0.90-0.98~$\mathrm{\mu}$m in this section.
Figure~\ref{fig_detectability_h2o} shows an intensity plot for the integration time required to detect the $\mathrm{H_2O}$ feature of 0.900-0.980~$\mathrm{\mu}$m with $\mathrm{S/N} = 5$ for three different Earth-trajectory epochs, 2.0~Ga~(a), 0.8~Ga~(b), and modern Earth~(c) with varying cloud altitude and coverage.
Same as the $\mathrm{O_2}$ case, for high altitude ($\gtrsim 3$~km for the modern case) clouds, a lower cloud coverage makes the feature deeper and the detection easier, while for low altitude ($\lesssim 3$~km for the modern case) clouds, a high cloud coverage makes the detection easier due to the increased flux at the continuum. 
\kawashima{While the water clouds also have absorption at this wavelength, the ice clouds do not have such absorption and thus their relatively reflective properties basically make the required integration time slightly smaller.}

Compared to the $\mathrm{O_2}$ case~(\S\ref{detec_o2}), the detection time for the $\mathrm{H_2O}$ feature significantly depends on the cloud properties, namely altitude and coverage.
Except for the extreme case of higher coverage ($\gtrsim 80$\% for the modern case) clouds at high altitudes ($\gtrsim 6$~km for the modern case), for all the three epochs, the detection of the $\mathrm{H_2O}$ feature will take approximately 3-10~hours, an order of magnitude smaller than that for the $\mathrm{O_2}$ feature. For the extreme cases, the minimum and maximum detection times are 0.4 to $6 \times 10^4$~hours for the 2.0~Ga case, 0.2 to $9 \times 10^3$~hours for the 0.8~Ga case, and 0.4 to $3 \times 10^5$~hours for the modern case (see Fig.~\ref{fig_detectability_h2o}). The very large detection times for the high altitude and high cloud coverage case is due to H$_2$O being less abundant in the upper atmosphere \kawashima{for planets with a cold trap}, whereas O$_2$ is well mixed.

{Note the water abundance for the two earlier geological epochs \kawashima{in our models} is largely determined by increased evaporation due to higher surface temperatures from a larger greenhouse effect despite a lower solar luminosity.

\begin{figure*}[ht!]
\gridline{\fig{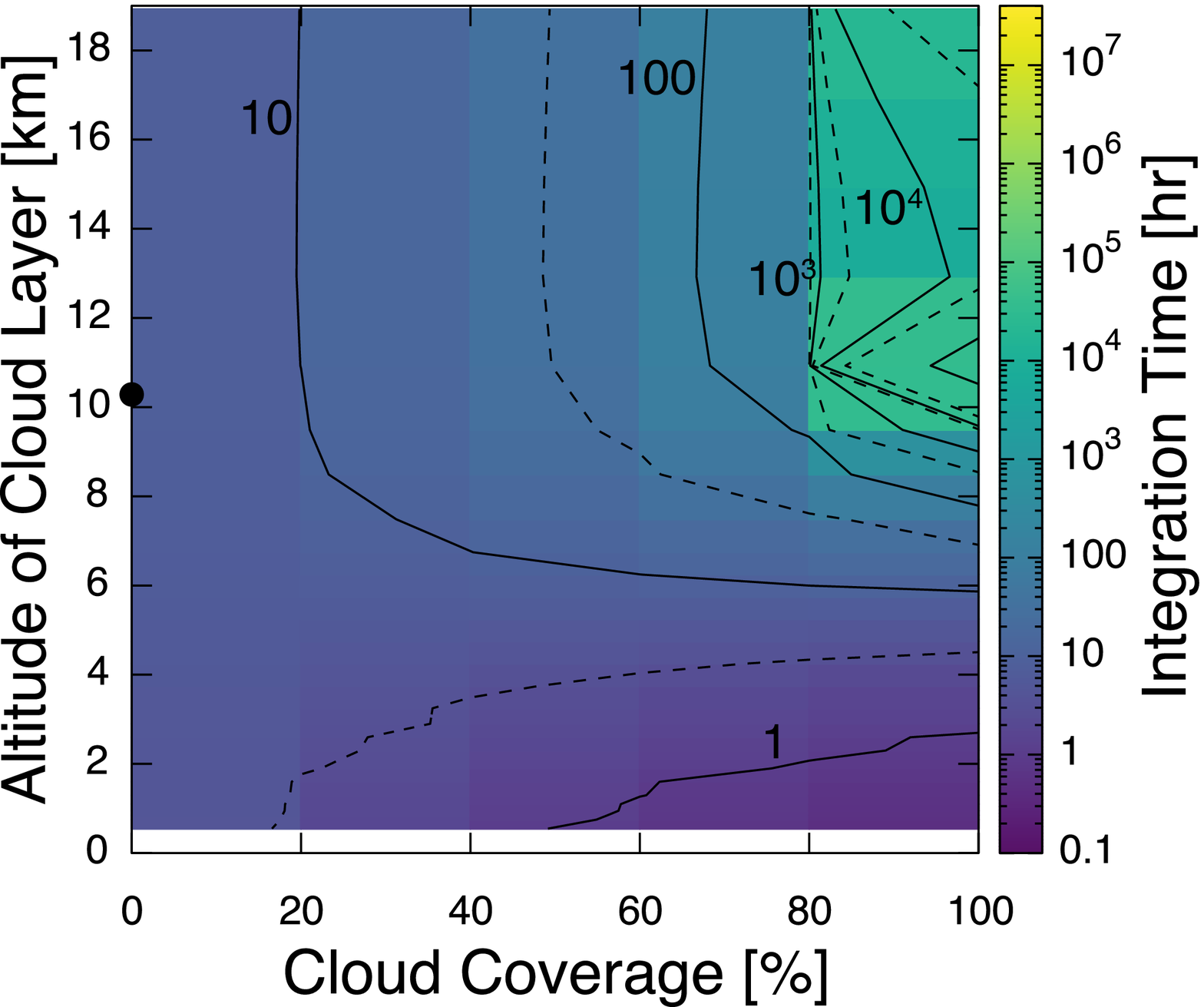}{0.4\textwidth}{(a)~2.0~Ga}
          \fig{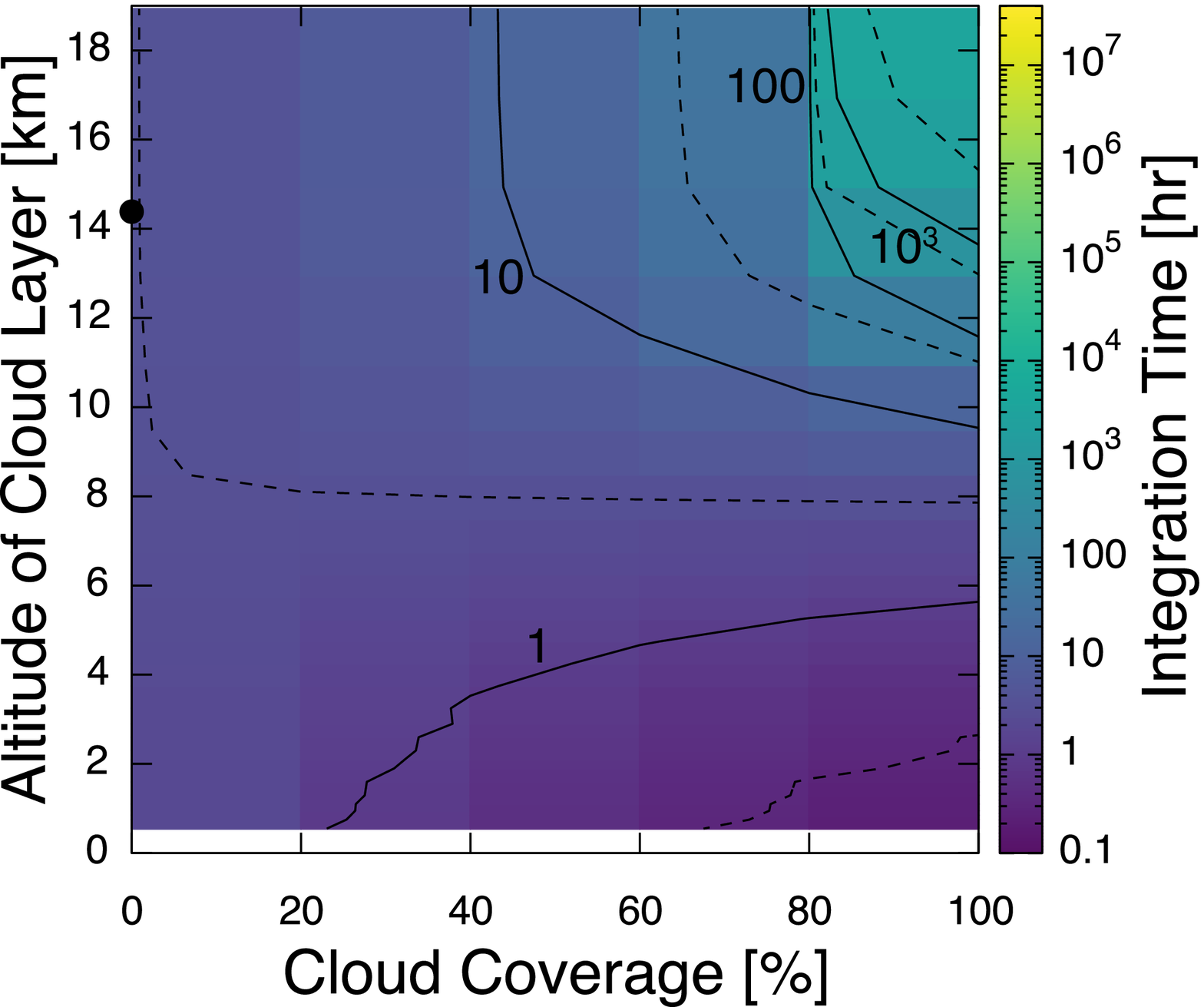}{0.4\textwidth}{(b)~0.8~Ga}
          }
\gridline{
          \fig{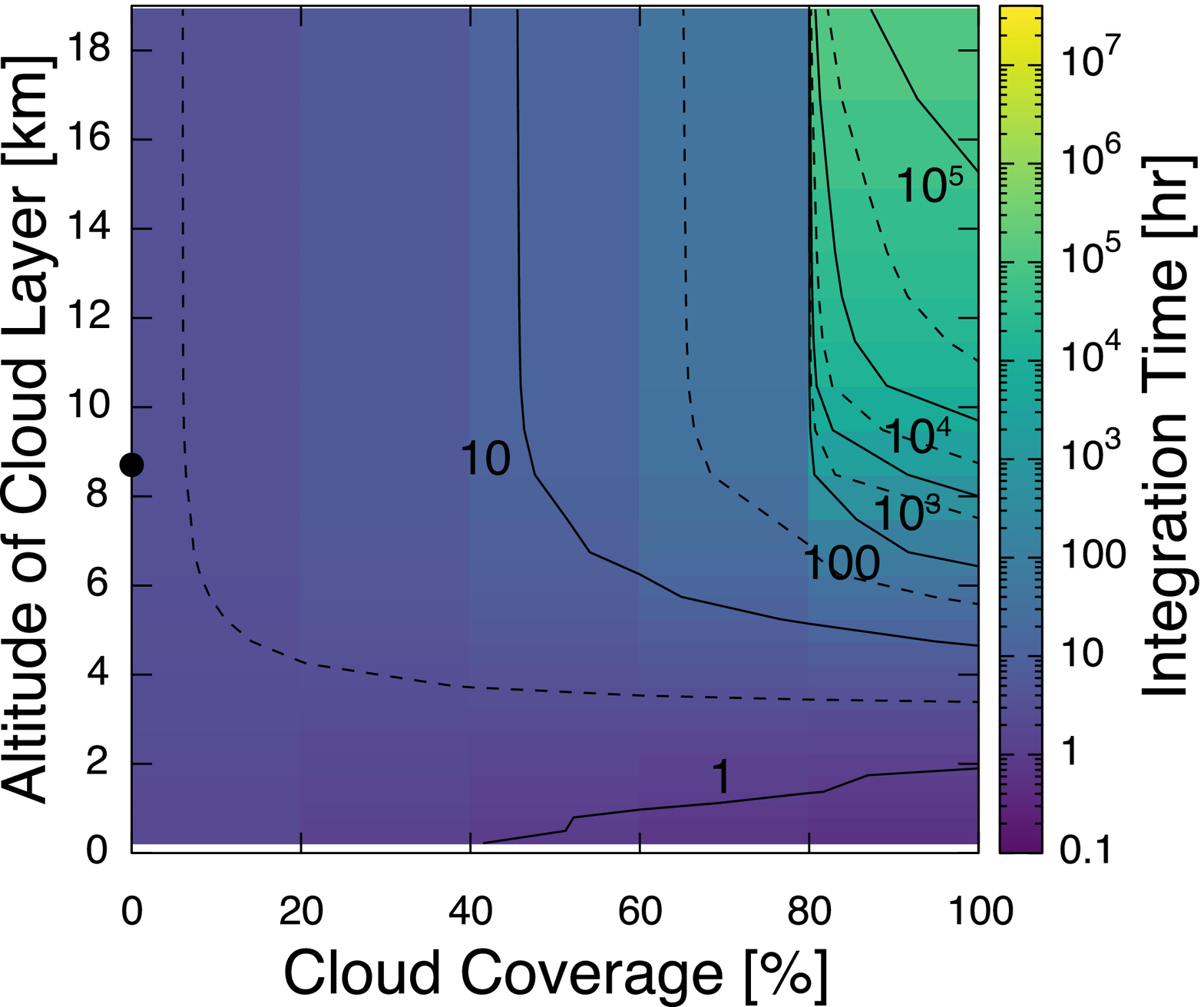}{0.4\textwidth}{(c)~Modern}
          }
\caption{Intensity plots for the integration time required to detect the $\mathrm{H_2O}$ feature of 0.90-0.98~$\mathrm{\mu}$m with $\mathrm{S/N} = 5$ are shown for three different Earth-like trajectory epochs, 2.0~Ga~(a), 0.8~Ga~(b), and the modern Earth~(c) with varying cloud altitude and coverage.
Contour lines for the integration time are also plotted in black solid and dashed lines. Note that the black dashed lines are for the integration times of 0.3, 3, 30, 300, $3 \times 10^3$, $3 \times 10^4$, and $3 \times 10^5$~hr but not labeled for the clarity. \kawashima{Also, the black filled circles on the vertical axes represent the altitude at 230~K, above which clouds are assumed as ice ones.} \label{fig_detectability_h2o}}
\end{figure*}

\subsection{$\mathrm{CH_4}$ Feature}
Figure~\ref{fig_detectability_ch4} shows an intensity plot for the integration time required to detect the strongest reflected light NIR $\mathrm{CH_4}$ feature at 1.64-1.78~$\mathrm{\mu}$m with $\mathrm{S/N} = 5$ for three different Earth-trajectory epochs, 2.0~Ga~(a), 0.8~Ga~(b), and modern Earth~(c) with varying cloud altitude and coverage.

\kawashima{Contrary to the cases of $\mathrm{O_2}$ and $\mathrm{H_2O}$, the optical properties of water and ice clouds are quite different in this wavelength region with much higher reflectivity for water clouds, and this causes the changes of the trend at the threshold altitudes.}
\kawashima{For the water cloud region at the lower altitudes,}
%\sout{the trend are the same as the cases of $\mathrm{O_2}$ and $\mathrm{H_2O}$ with low altitude clouds:} 
the lower the altitude of the cloud layer becomes, and the higher the cloud coverage becomes, the detection time becomes smaller. \kawashima{This is because of the larger column-integrated concentration of the species above the cloud layer and the relatively reflective properties of water clouds compared to the surface}.
\kawashima{However, in the ice cloud region at the higher altitudes, the lower the altitude of the cloud layer becomes, and the 
lower the cloud coverage becomes, the detection time becomes smaller. This is due to the larger column-integrated concentration of the species above the cloud layer and the relatively absorbing properties of ice clouds compared to the surface in this wavelength range.}

The detection of the $\mathrm{CH_4}$ feature will take approximately 10 and 30~hours for 2.0~Ga and 0.8~Ga cases, respectively.
For the extreme cloud parameter cases, the minimum and maximum detection times are 1 to 300~hours for the 2.0~Ga case and 2 to 900~hours for the 0.8~Ga case.
\kawashima{Here we note that the $\mathrm{CH_4}$ abundances for 2.0~Ga and 0.8~Ga cases are not well-constrained and the values we adopt may be optimistic \citep[e.g.,][]{2017AsBio..17..287R}.}

For the modern Earth case, however, it will take more than 6000~hours for all the cloud parameters modeled and the feature is undetectable even with the LUVOIR-sized telescope regardless of the cloud parameters because of the relatively low $\mathrm{CH_4}$ abundance in the modern atmosphere and the weaker NIR feature as compared with the IR CH$_4$ feature. For the extreme cases, the minimum and maximum detection times are 6000 to $3 \times 10^7$~hours for the modern case (see Fig.~\ref{fig_detectability_ch4}).
The trend for the modern Earth case is the same as the \kawashima{2.0~Ga and 0.8~Ga cases}.
%\sout{$\mathrm{O_2}$ and $\mathrm{H_2O}$ cases: for high ($\gtrsim 13$~km) altitude clouds, a lower cloud coverage makes the feature deeper and the detection easier while for low ($\lesssim 13$~km) altitude clouds, a high cloud coverage makes the detection easier due to the increased flux at the continuum}.

\begin{figure*}[ht!]
\gridline{\fig{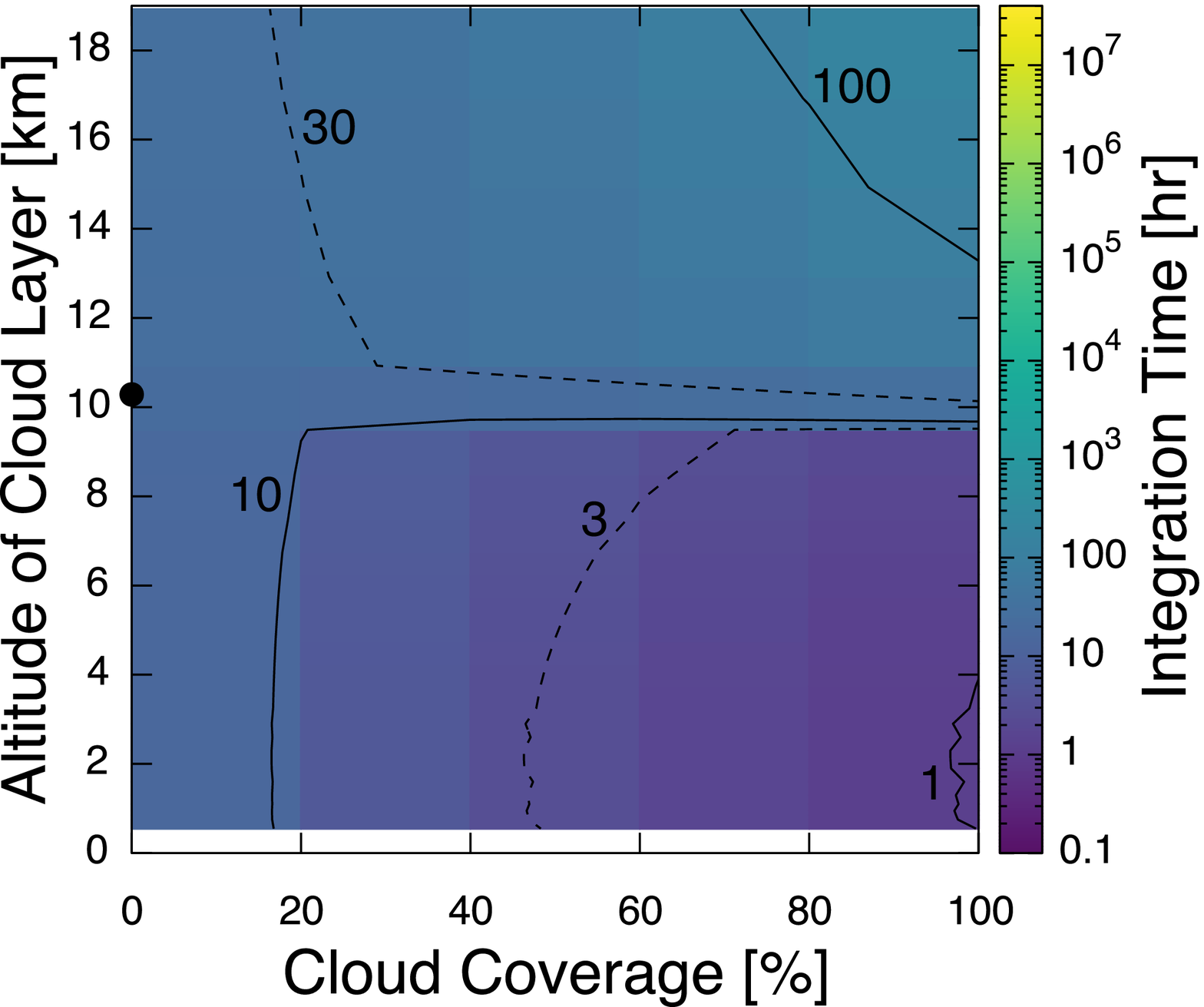}{0.4\textwidth}{(a) 2.0 Ga}
          \fig{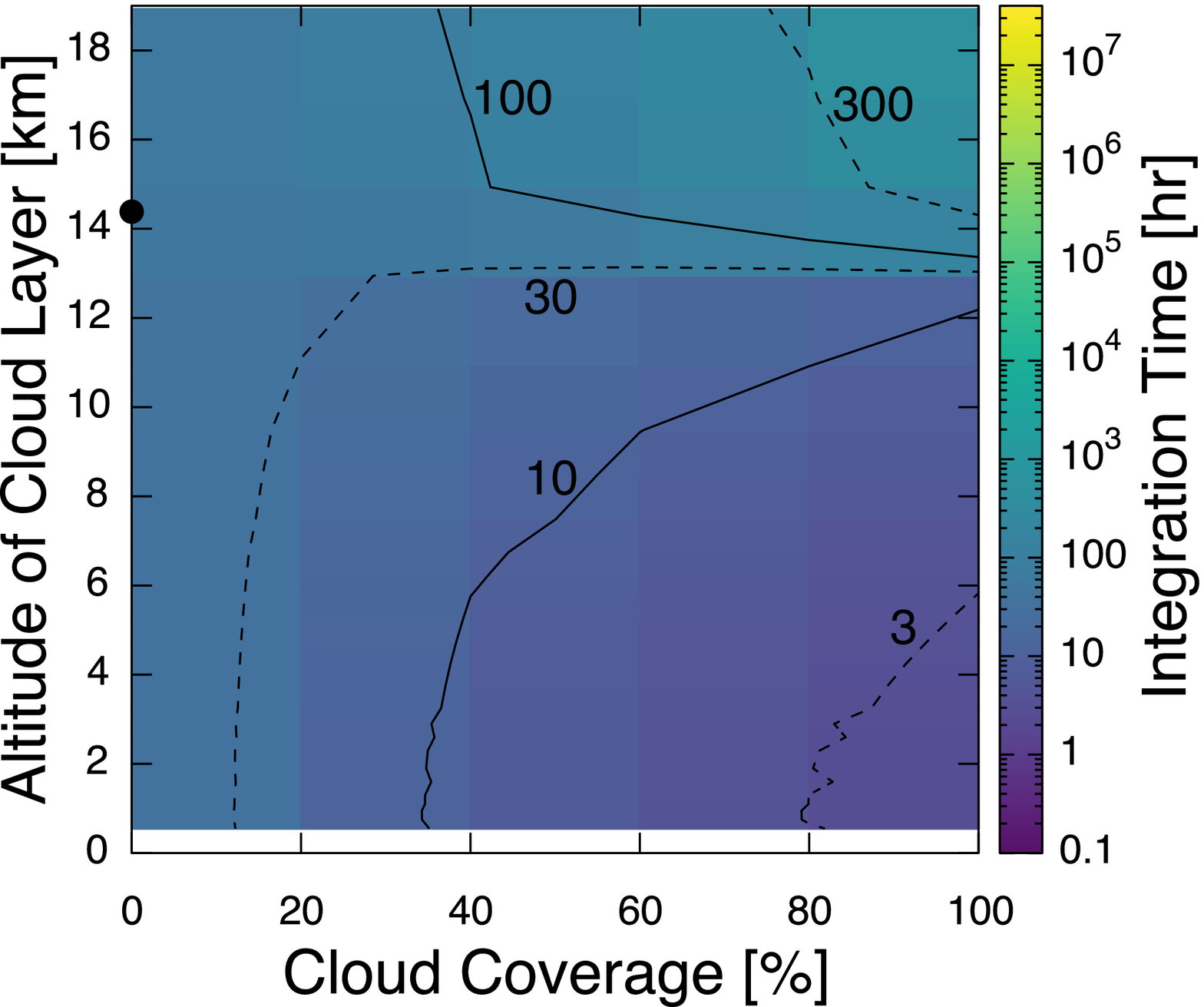}{0.4\textwidth}{(b) 0.8 Ga}
          }
\gridline{
          \fig{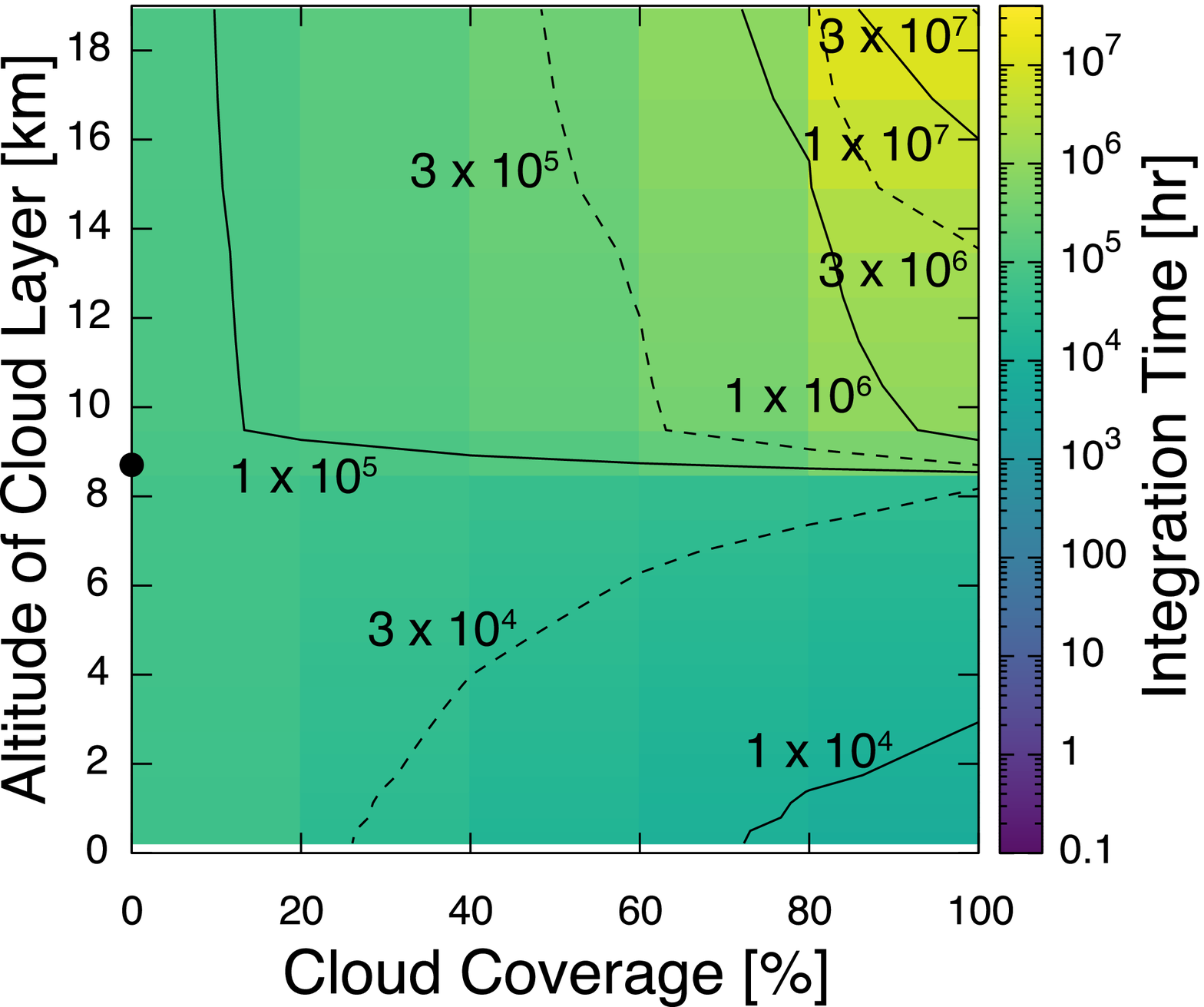}{0.4\textwidth}{(c)~Modern}
          }
\caption{Intensity plots for the integration time required to detect the $\mathrm{CH_4}$ feature of 1.64-1.78~$\mathrm{\mu}$m with $\mathrm{S/N} = 5$ are shown for three different Earth-like trajectory epochs, 2.0~Ga~(a), 0.8~Ga~(b), and the modern Earth~(c) with varying cloud altitude and coverage. Contour lines for the integration time are also plotted in black solid and dashed lines. \kawashima{Also, the black filled circles on the vertical axes represent the altitude at 230~K, above which clouds are assumed as ice ones.}
\label{fig_detectability_ch4}}
\end{figure*}

\section{Discussion} \label{sec:disc}
In this study, we have examined the impact of cloud properties (cloud altitude and its coverage) on the detectability of the molecules on the reflectance spectra of an Earth-like planet at different geological epochs systematically. A self-consistent microphysical model \kawashima{\citep[e.g.,][]{2012Icar..221..603Z, 2017ApJ...835..261O, 2018ApJ...859...34O, 2018ApJ...860...18P, 2018ApJ...863..165G, 2019A&A...622A.121O}} would be needed to examine the plausibility of these cloud parameters, which is beyond the scope of this study. In our two-stream radiative transfer model, the cloud layer is assumed to be completely \kawashima{absorbing or} reflective surface. In reality, however, some light can penetrate the cloud layer depending on the thickness of the cloud layer, and the cloud particle size. This effect also cannot be studied without deriving the distributions of the size and number density of the cloud particles by a microphysical cloud model and by using a multi-scattering radiative transfer model.

For \kawashima{all the cases of} detecting the $\mathrm{O_2}$ A-band, \kawashima{$\mathrm{H_2O}$ feature at 0.90-0.98~$\mathrm{\mu}$m, and that of $\mathrm{CH_4}$ at 1.64-1.78~$\mathrm{\mu}$m,} we have found that the shortest integration times are for a high-coverage, low-altitude cloud layer due to the deeper absorption created with increased back-scattered light of the higher albedo cloud layer when compared with the surface albedo and the larger integrated column density. It is possible that this same effect on the detectability could be seen on a snowball planet \kawashima{\citep[e.g.,][]{2008ApJ...680L..53T, 2014ApJ...790..107K, 2015ApJ...815L...7K, 2016ApJ...825L..21K}}. While the colder temperatures may slow down bioproductivity, Earth has been through several snowball states with global glaciation well after oxygen has been a major atmospheric constituent and even quite recent in its history \citep[see][and references therein]{kirschvink1992late, hoffman2017snowball}. These snowball states may make these features easier to detect \kawashima{as long as there are} appreciable levels of \kawashima{these species} in the atmosphere.
\kawashima{As for the abundance of $\mathrm{H_2O}$, although we have assumed larger abundances for the two earlier geological epochs, during the cooler period within the huge temporal range of temperature oscillation, it might be lower than what we have assumed, making the detection more difficult.}

\kawashima{While we explored the detectability of specific absorption features of $\mathrm{O_2}$, $\mathrm{H_2O}$, and $\mathrm{CH_4}$ via the low-resolution measurement of the flux-contrast around the wavelength range of absorption features by LUVOIR, \cite{2018JATIS...4c5001W} investigated the detection time of $\mathrm{O_2}$, $\mathrm{H_2O}$, $\mathrm{CH_4}$, and $\mathrm{CO_2}$ via high-resolution cross-correlation technique over the wavelength range of $0.5-1.8$~$\mu$m by HabEx and LUVOIR.
In their LUVOIR case, considering an modern Earth-like planet with a clear sky atmosphere located at 5~pc away, they reported that the required starlight suppression for an exposure time of 100~hours are $\sim 2 \times 10^{-9}$ and $\sim 10^{-8}$ for $\mathrm{H_2O}$ and $\mathrm{O_2}$, respectively, while $\mathrm{CH_4}$ and $\mathrm{CO_2}$ are undetectable with 100-hour exposure time.
}
\section{Summary} \label{sec:summary}
In this study, we have explored the effect of cloud altitude and its coverage on the reflectance spectra of Earth-like planets at different geological epochs and examined the detectability of astrobiologically interesting gaseous molecules in the visible and near-infrared spectrum, namely $\mathrm{O_2}$, $\mathrm{H_2O}$, and $\mathrm{CH_4}$, by simulating instrumental noise for the proposed mission concept LUVOIR.

Considering an Earth-like planet located at 5~pc away, we have found that for the proposed LUVOIR telescope, the detection of the $\mathrm{O_2}$ A-band feature (0.76~$\mathrm{\mu}$m) will take approximately 100, 30, and 10~hours for the majority of the cloud parameters modeled for atmospheres with 0.1, 0.5, and 1.0~PAL O$_2$ abundances, respectively.
Especially, for 0.5 and 1.0~PAL $\mathrm{O_2}$ cases, the feature could be detectable with integration times $\lesssim 10$~hours as long as there are lower altitude ($\lesssim 8$~km) clouds with a global coverage of $\gtrsim 20\%$.
For 0.01~PAL O$_2$ case, however, it will take more than 100~hours for all the cloud parameters modeled.

The combined detection of O$_2$ and CH$_4$ remains the strongest biosignature. There is currently no known abiotic oxygen production mechanism which would persist with a simultaneously dedectable amount of CH$_4$ present. For $\mathrm{CH_4}$, we have found that the detection of its NIR feature at 1.64-1.78~$\mathrm{\mu}$m will take approximately 10 and 30~hours for 2.0~Ga and 0.8~Ga cases, respectively.

For the modern Earth case, however, it will take more than 6000~hours for all the cloud parameters modeled and the feature is undetectable even with the LUVOIR-sized telescope because of the relatively low $\mathrm{CH_4}$ abundance in the modern atmosphere.

While H$_2$O is not a biosignature, it is an important indicator of habitability and provides necessary context for interpreting future exoplanet observations. For $\mathrm{H_2O}$ feature at 0.90-0.98~$\mathrm{\mu}$m, we have found that except for the extreme case of higher cloud coverage at high altitudes, the detection of its strongest feature will take approximately 3-10 hours.

In summary, a LUVOIR-sized mission with a coronagraph could detect the reflected light of O$_2$, CH$_4$, and H$_2$O for many cases comparable to Earth's geological history with a wide range of cloud parameters. To detect the combination of these gases with less than 100 hours of observation time, however, will require more CH$_4$ than in modern Earth's atmosphere, O$_2$ levels around 0.1~PAL or greater, and clouds that are lower in altitude or patchy in coverage.

\acknowledgments

We thank Tyler D. Robinson for kindly providing his noise calculation model and giving us advice and comments.
\kawashima{We are grateful to Mark Claire for providing helpful comments on the geological constraints on the past oxygen and methane abundances.}
We would like to thank the Kavli Summer Program in Astrophysics 2016; Exoplanetary Atmospheres for providing us the opportunity to conduct this research and Kavli Foundation for supporting the program.
\kawashima{We also thank the anonymous referee for his/her careful reading and constructive comments, which helped us improve this paper greatly.}
Y.K. is supported by the Grant-in-Aid for JSPS Fellow (JSPS KAKENHI No.15J08463), Leading Graduate Course for Frontiers of Mathematical Sciences and Physics, Grant-in-Aid for Scientific Research (A) (JSPS KAKENHI No.15H02065), and the European Union's Horizon 2020 Research and Innovation Programme under Grant Agreement 776403.
This work was also supported by a grant from the Simons Foundation (SCOL award 339489 to S.R.) and benefited from the Exoplanet Summer Program in the Other 
Worlds Laboratory (OWL) at the University of California, Santa Cruz, a 
program funded by the Heising-Simons Foundation.

\bibliography{ads}



\end{document}